\documentclass{aa} 

\usepackage{graphicx}
\usepackage{txfonts}
\usepackage{subcaption}
\usepackage{xcolor}
\usepackage{multirow}
\usepackage{hyperref}
\usepackage{placeins}
\usepackage{stfloats}
\usepackage{flushend}

\def\flx{erg~cm$^{-2}$~s$^{-1}$}
\def\lum{erg~s$^{-1}$}

\def\arcsec{\hbox{$^{\prime\prime}$}}

\begin{document}

   \title{Optical observations of candidate host galaxies of eight fast X-ray transients}

   \subtitle{}
    \author{Agnes P. C. van Hoof
          \inst{1},
          Peter G. Jonker\inst{1},
          Lieke Tommel\inst{1},
          Jonathan A. Quirola-V\'asquez\inst{1},
          Daniel Mata S\'anchez\inst{2}\fnmsep\inst{3},
          Joyce N. D. van Dalen\inst{1},
          Andrew J. Levan\inst{1}\fnmsep\inst{4},
          Morgan Fraser\inst{5},
          Ann Zabludoff\inst{6},
          Manuel A. P. Torres\inst{2}\fnmsep\inst{3},
          Javi S\'anchez-Sierras\inst{1},
          Antonio Martin-Carrillo\inst{5},
          Thomas Wevers\inst{7},
          Marco Berton\inst{8},
          Minghao Yue\inst{9},
          Vik S. Dhillon\inst{10}\fnmsep\inst{2},
          Franz E. Bauer\inst{11},
          Stuart P. Littlefair\inst{12}
          }

   \institute{Department of Astrophysics/IMAPP, Radboud University, 6525 AJ Nijmegen, The Netherlands
    \and Instituto de Astrof\'isica de Canarias, E-38205 La Laguna, Tenerife, Spain  
    \and Departamento de Astrofísica, Univ. de La Laguna, E-38206 La Laguna, Tenerife, Spain 
    \and Department of Physics, University of Warwick, Coventry, CV4 7AL, UK
   \and School of Physics and Centre for Space Research, University College Dublin, Belfield, Dublin 4, Ireland
   \and Steward Observatory, University of Arizona, 933 North Cherry Avenue, Tucson, AZ 85721-0065, USA
   \and Astrophysics \& Space Center, Schmidt Sciences, New York, NY 10011, USA
   \and European Southern Observatory, Alonso de C\'ordova 3107, Vitacura, Santiago, Chile
   \and  MIT Kavli Institute for Astrophysics and Space Research, 77 Massachusetts Ave., Cambridge, MA 02139, USA
   \and Astrophysics Research Cluster, School of Mathematical and Physical Sciences, University of Sheffield, Sheffield S3 7RH, UK
   \and Instituto de Alta Investigaci\'on, Universidad de Tarapac\'{a}, Casilla 7D, Arica, Chile
   \and Department of Physics and Astronomy, University of Sheffield, Sheffield, S3 7RH, United Kingdom  
   }

   \date{Received , ; accepted , }

  \abstract
   {Fast X-ray transients (FXTs) are extragalactic flashes of X-rays with a typical duration of minutes to hours for which a variety of origins has been proposed and observed.
}
   {To decipher the origin of FXTs, particularly those lacking multi-wavelength counterparts, we aim to understand their energetics and environments.}
   {We present deep optical ground-based observations of the positions of eight FXTs in order to try and identify and characterize candidate host galaxies. We use their properties to discriminate between possible progenitor scenarios.
   }
   {We identify candidate host galaxies for Swift~J050400.2+673405, XRT140507, XRT040610, XRT151121, XRT191127 and EP240708a. For each candidate, we infer the spectroscopic or photometric redshift, stellar mass, star formation rate, metallicity and stellar population age by fitting our data with the spectral energy distribution fitting code BAGPIPES. 
   We re-identify XRT191223 as a Galactic stellar flare. 
    }
   {For several FXTs, there are multiple candidate host galaxies, which complicates deriving constraints on the origin of the FXT. 
   Assuming association with (one of) those candidates, all are consistent with a (non-)relativistic white dwarf -- intermediate mass black hole tidal disruption event (WD-IMBH TDE) and a binary neutron star (BNS) merger. Two are consistent with a supernova shock breakout and only EP240708a with cocoon emission from a long gamma-ray burst. We also discuss the possibility that the host galaxies remain undetected in our observations.
   We conclude that FXTs detected by \emph{Chandra} and XMM-\emph{Newton} are likely to arise from a variety of origins, and we discuss that part of this population differs from FXTs detected by Einstein Probe many of which appear consistent with a collapsar scenario.
   }


\titlerunning{Candidate host galaxies of eight FXTs}
\authorrunning{A.P.C. van Hoof et al.}
   \maketitle
 \nolinenumbers

\section{Introduction}
\label{SectIntroduction}
Fast X-ray transients (FXTs) are singular flashes of soft X-ray photons, with energies between 0.3 -- 10 keV, lasting minutes to hours. They originate from extragalactic phenomena. 
About 30 FXTs have been identified through extensive searches in the \emph{Chandra} \& XMM-\emph{Newton} archives \citep{Jonker_2013, Glennie2015, Irwin_2016, Bauer_2017, Xue_2019, AlpLarsson2020, Novara_2020, Lin_2022, Quirola_Vasquez_2022, Quirola_Vasquez_2023}. One has been detected serendipitously with the X-ray telescope on the Swift observatory \citep[e.g.~][]{Soderberg_2008}, and the Swift team has also been reporting serendipitously discovered FXTs (\cite{2023MNRAS.518..174E}). Currently, the X-ray satellite Einstein Probe \cite[EP;][]{yuan2015einstein,Yuan_2022}), specifically designed to detect X-ray transients and rapidly notify the scientific community about their discovery, is revolutionizing the field by discovering about a hundred FXTs per year. The Chandra \& XMM-Newton FXTs have been discovered retrospectively, long after the outburst, which made timely follow-up and multi-wavelength counterpart identification impossible. 
It is only since the launch of EP that more FXTs are being reported in close to real time and multi-wavelength counterparts can be identified. Nonetheless, despite the rapid response opportunities, for about $\sim$45\% of the EP-discovered FXTs during the first year of operations no multi-wavelength counterpart was discovered.
This lack of counterparts leaves their origins largely unconstrained. Fortunately, studying the environments of FXTs can reveal important information constraining their progenitor mechanisms. 

Various possible origins to explain FXTs have been proposed. 
Since the detection of FXTs with EP, the association of FXTs with collapsar (or as previously called, long duration) gamma-ray bursts (LGRBs) has been revealed by the appearance of evidence for a broad-lined type Ic (Ic-BL) SN in the optical follow-up data \citep[e.g.,][]{vandalen2025, Srivastav2025, rastinejad2025, quirolavasquez2026, vanhoof2026}. Some of these FXTs coincided with a $\gamma$-ray detection \citep[e.g.,][]{levan2025, jiang2025}, while others did not \citep[e.g.,][]{2024GCN.38435....1R, sun2025, Eyles-Ferris2025}. 
One explanation for the lack of $\gamma$-rays for some of these FXTs is that our line-of-sight lies outside the jet-core.
However, \cite{Wichern2024} showed that it is unlikely that all \emph{Chandra}-discovered FXTs can be explained as GRBs observed off-axis. A second explanation for the detection of a bright X-ray signal in an FXT without the detection of a bright gamma-ray signal, is provided by a so-called ``dirty fireball". After the collapse of the massive star, a fireball, an ultra-relativistic outflow, is produced. However, when the outflow is ``dirty" it contains more baryonic matter than when it is not "dirty". The addition of baryons produces a lower Lorentz factor ($<<$ 100) jet. The peak emission shifts to lower energies, into the X-ray regime and the $\gamma$-ray luminosity is too low to be detected by current instrumentation \citep{paczynski1998, Dermer1999}. However, the rate of these events cannot be significantly higher than the rates of classical GRBs due to limits based on optical detections of these dirty fireballs \citep{Cenko2013, Ho2018, Ho2020}.
The collapsar scenario also gives rise to a cocoon due to the jet propagation within the stellar envelope. The jet injects part of its energy into the surrounding material also spreading sideways, forming a cocoon \citep{Nakar2017, Izzo2019}. This radiation of the hot cocoon is predicted to be mostly thermal and to peak in X-rays \citep{DeColle2018}. The X-rays can reach peak luminosities of L$_{X,peak}$ \textgreater 10$^{47}$ erg s$^{-1}$. 
In the case of a dirty fireball, the jet is choked and only the cocoon will break out with peak X-ray luminosities of L$_{X,peak}$ $\approx$ 10$^{45}$ - 10$^{47}$ erg s$^{-1}$.
The majority of LGRBs are the result of massive star collapse, and are therefore occurring in star-forming regions. LGRBs and dirty fireballs predominantly occur in low-metallicity environments \citep{Kouveliotou2012}, which for nearby sources ($z \leq 0.7$) are most likely to be dwarf galaxies \citep{Tremonti2004,Zhou2024}. 

Before EP was launched, other models for FXTs found in archives were proposed. One of these models involves a binary neutron star (BNS) merger that leads to the formation of a highly magnetized, millisecond spin period neutron star (i.e.~a magnetar). Its spin-down energy powers the X-ray signal \citep{Metzger2008,Rowlinson_2013, lu2014, Siegel2016}.
BNS mergers can occur with a large offset to the host galaxy and are predicted to have peak luminosities of L$_{X,peak}$ $\approx$ 10$^{44}$ - 10$^{47}$ erg s$^{-1}$ \citep[e.g.][]{Dai2006,Metzger2008,Yu2013,Sun2017, quirolavasquez2024}.

Another candidate progenitor is that where an intermediate-mass black hole (IMBH) tidally disrupts a white dwarf (WD). The X-ray luminosity is powered by the accretion of stellar debris onto the IMBH, which launches a relativistic jet. If so, the X-ray luminosity could reach peak luminosities of L$_{X,peak}$~$\lesssim$~10$^{43}$~erg~s$^{-1}$ when there is no relativistic emission, and become as high as 10$^{50}$~erg~s$^{-1}$ when there is relativistic emission directed towards us \citep{Maguire_2020}. Such tidal disruption events (TDEs) produce light curves with typical timescales of seconds to hours, making them good candidates for the progenitors of some FXTs. The expected rate for these events is 10–300 yr$^{-1}$ Gpc$^{-3}$ \citep{Maguire_2020}, (the large range reflects uncertainties in the volumetric density of IMBHs), while the detected FXTs in Chandra and EP suggest a rate of $\sim$10$^{4}$ FXTs yr$^{-1}$ Gpc$^{-3}$ \citep{Quirola_Vasquez_2023}. Therefore, this progenitor mechanism is unable to explain the nature of all FXTs.
About 90\% of TDEs involving IMBHs are thought to reside in globular clusters (GCs) within a maximum radius of $\sim$2~-~15~kpc away from the host's galactic nucleus \citep{Fragione_2018, Beckmann_2023}. 

The shock that crosses the surface of an exploding star, called the supernova (SN) shock breakout (SBO), is for compact massive stars (such as Wolf-Rayet stars) initially visible in X-rays \citep{Soderberg_2008, Waxman_2017}. The emission of these SBOs evolves into UV and optical as the envelope cools, which can be observed on timescales longer than one day. On the other hand, core-collapse red supergiants produce much slower and cooler SBOs and no X-rays. 
Any X-ray emission from SBOs typically lasts only from tens of seconds to minutes and has a peak luminosity of L$_{X,peak}$ $\approx$ 10$^{42}$ - 10$^{44}$ erg s$^{-1}$ \citep[e.g.][]{Ofek_2014}.
A related model to this is the phantom SBO \citep{Paradiso2024}. Here, the shock wave generated by the collapse of the stellar core is similar to the binding energy of the star. The shock decelerates to a speed comparable to the escape velocity of the progenitor. The shock stalls within the progenitor, resulting in a failed supernova, or stalls close to the surface. When the outer-envelope material of the star then falls through the stalled shock, an SBO-like signal is expected but without an SN. This yields an FXT, but without a subsequent optical SN signature.

The peak luminosity in X-rays and the host galaxy properties such as star formation rate (SFR), galaxy mass, metallicity and offset varies for the different proposed origins of FXTs. Hence, they are a powerful diagnostic for constraining the progenitor to individual FXTs \citep[e.g.,][]{AlpLarsson2020, Lin2022,Quirola_Vasquez_2022, Quirola_Vasquez_2023, Eappachen_2022, Eappachen_2023, Inkenhaag_2024}. These previous studies showed that the distributions of these FXT host parameters is wide, suggesting that FXTs have multiple progenitor mechanisms. This is supported by the diversity in X-ray properties of FXTs \citep[e.g.][]{Quirola_Vasquez_2022,Quirola_Vasquez_2023}.

In this work, we present deep optical observations designed to find and characterize the host galaxies of eight FXTs detected by \emph{Chandra}, \emph{XMM Newton}, \emph{Swift}, and EP. We aim to constrain the nature of these FXTs based on the host properties. This paper is organized as follows. We describe the sample of FXTs in the Appendix~\ref{SectSample}. The observations are described in Section~\ref{SectObservations}. We present the results in Section~\ref{SectResults} and discuss them in Section~\ref{SectDiscussion}. We end with our conclusions in Section~\ref{SectConclusion}. 
Throughout this work, all magnitudes are given in the AB magnitude system. Uncertainties are given at 1$\sigma$, unless otherwise mentioned. We assume a flat Lambda cold dark matter ($\Lambda$CDM) Planck cosmology \citep{Planck2020} with H$_0$=67.7 km s$^{-1}$ Mpc $^{-1}$ and $\Omega_{m}$=0.31 throughout this work.

\section{Sample of FXTs}
\label{SectSample}
Here, we describe where the eight FXTs that are studied in this work are discovered and, if available, some information of previously reported candidate host galaxies. We provide a short summary table in Appendix \ref{AppendixSample}. XRT040610 and XRT100424 were found by \cite{AlpLarsson2020} in the XMM-\emph{Newton} archive, while XRT140507, XRT151121, XRT191127 and XRT181223 were found by \cite{Quirola_Vasquez_2023} in the \emph{Chandra} archive. For XRT040610, XRT140507 and XRT151121, an (extreme) Galactic stellar flare origin could not be ruled out in the literature.

For XRT040610, there is no underlying source in the uncertainty region of this FXT in Pan-STARRS or 2MASS. However, \cite{AlpLarsson2020} propose the redshift to be equal to the average photometric redshift of eight neighbouring galaxies in SDSS12 of $z=0.5 \pm 0.17$.

XRT100424 lies in projection close to an extended object with an offset of 0.5$\pm$1.7\arcsec. \cite{AlpLarsson2020} suggest this is the host galaxy of this FXT and that it has a redshift in the range of 0.08–0.4. The FXT has a peak luminosity of L$_{X,peak} = (0.05^{+0.12}_{-0.01}) \times 10^{44}$~erg~s$^{-1}$ at $z=0.13$, consistent with an SN SBO. 

For XRT140507, no underlying galaxies were found in archival images of DECam, HST, 2MASS, and unWISE. The deepest limiting apparent magnitude is $m_g > 24$.

For XRT151121, no optical and NIR sources were detected within the 3$\sigma$ X-ray localization region in Pan-STARRS, 2MASS, and unWISE. The deepest limiting apparent magnitude is $m_g > 24.1$.

XRT191127 lies close (offset $\sim$0.6") to a faint source with $m_r \approx 23.5$ AB mag detected in DECam $r^\prime$-band images. This source is considered to be the host galaxy, but no redshift is known. 

XRT191223 has an offset of 0.5$\pm$0.2\arcsec to its faint ($m_r\approx 25.1$) potential host detected in Pan-STARRS. There are no detections of this candidate host in 2MASS NIR or unWISE IR images.

Swift~J050400.2+673405 (Swift~J050400 hereafter) was detected by the Swift satellite \citep{Gehrels2004} and presented in the Living Swift XRT Point Source catalogue (LSXPS) from \cite{Evans2023}\footnote{\url{https://www.swift.ac.uk/LSXPS/transients/3435}}. This X-ray transient has not been reported in the literature before.

EP240708a was detected by the Wide-field X-ray Telescope (EP-WXT) on board the Einstein Probe (EP) satellite \citep{2024GCNEP240708a} and was four hours later also detected by the Fast X-ray Telescope on EP (EP-FXT) \citep{2024GCNFXTEP240708a}. No optical counterpart was detected for this FXT, with the deepest 3$\sigma$ upper limit $r>24.1$ from the Nordic Optical Telescope (NOT) at $\sim$4 hours after the EP-WXT trigger \citep{EP240708aGCN2024NOT}.

\section{Observations}
\label{SectObservations}
We have obtained deep broadband images and long-slit spectra of  potential host galaxies located within 3$\sigma$ of the position of each FXT using the Gran Telescopio Canarias (GTC), Magellan, Very Large Telescope (VLT), Southern Astrophysical Research Telescope (SOAR) and Gemini South. A journal of the photometric and spectroscopic observations is given in Table~\ref{table:obsjournal}. Additionally, we have obtained \textit{g, r, i, z} images of the host of XRT191223 from the Subaru Hyper Suprime-Cam (HSC) archive.  

All images taken with FORS2 with seeing larger than 0.9\arcsec~ were excluded, as there are many observations with excellent seeing. Data obtained on nights when the extinction due to the earth's atmosphere, as determined from the night logs, was more than 1$\sigma$ above the mean of the observations for that source were excluded. Those images that remained after this filtering, were averaged. For a few nights no extinction values were reported, for these cases we checked if adding those images to the average resulted in a deeper image. These two filter criteria combined, resulted in the exclusion of the observations carried out during the night of 2023 June 28 and 29.

All images were bias and flat field corrected using standard tasks in \texttt{PyRAF} \citep{Pyraf2012} or with the dedicated pipeline for HiPERCAM \citep{2021Dhillon}  and the FORS2 workflow in ESOReflex \citep{ESOReflex}. Images within a night were average combined using \texttt{imcombine} in IRAF \citep{iraf}. Calibration to the world coordinate system  was performed either with \url{astrometry.net} \citep{astrometrynet} or the Graphical Astronomy and Image Analysis Tool \citep[GAIA][]{GAIA_Draper2014}, reaching in all cases an astrometric solution with root mean square $<<1$\arcsec. A more detailed description of the instruments used and the data reduction of the spectra is given in Appendix~\ref{app:obs+datared}.

\section{Results}
\label{SectResults}
\subsection{Candidate host galaxy identification}
We use the Source Extractor software \citep[SE;][]{sextractor} to determine the RA and Dec of all the objects located within the region corresponding to the 2$\sigma$ uncertainty on the position of each of the FXTs. Extended objects that fall inside such a region are considered a candidate host galaxy to the FXT. In some cases, we also consider a bright galaxy even if it falls outside this region to be a candidate host galaxy of the FXT, if the chance alignment probability \citep[P$_{chance}$][]{Bloom2002} is similar to that of a candidate host galaxy within the region defined above. The coordinates for the newly found host galaxies are given in Table~\ref{table:hostsall}. We show finder charts for the candidate hosts of each FXT in Fig.~\ref{fig:finders}. An exception is XRT100424 for which we show the finder chart in Appendix~\ref{Section:AppendixFinders} as its host was discovered before \citep{AlpLarsson2020}. Below, we describe the host galaxy identification in detail per FXT.

\begin{figure*}[ht!]
\sidecaption
\begin{minipage}[]{0.49\textwidth}
\includegraphics[width=\linewidth]{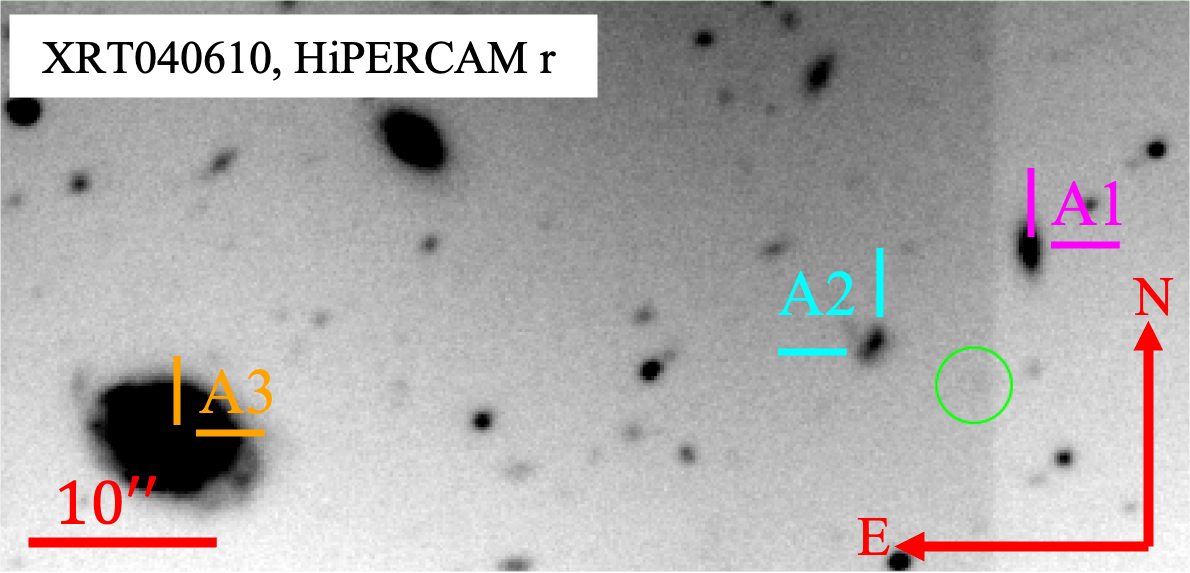}\vspace{0.1cm}
  \includegraphics[width=\linewidth]{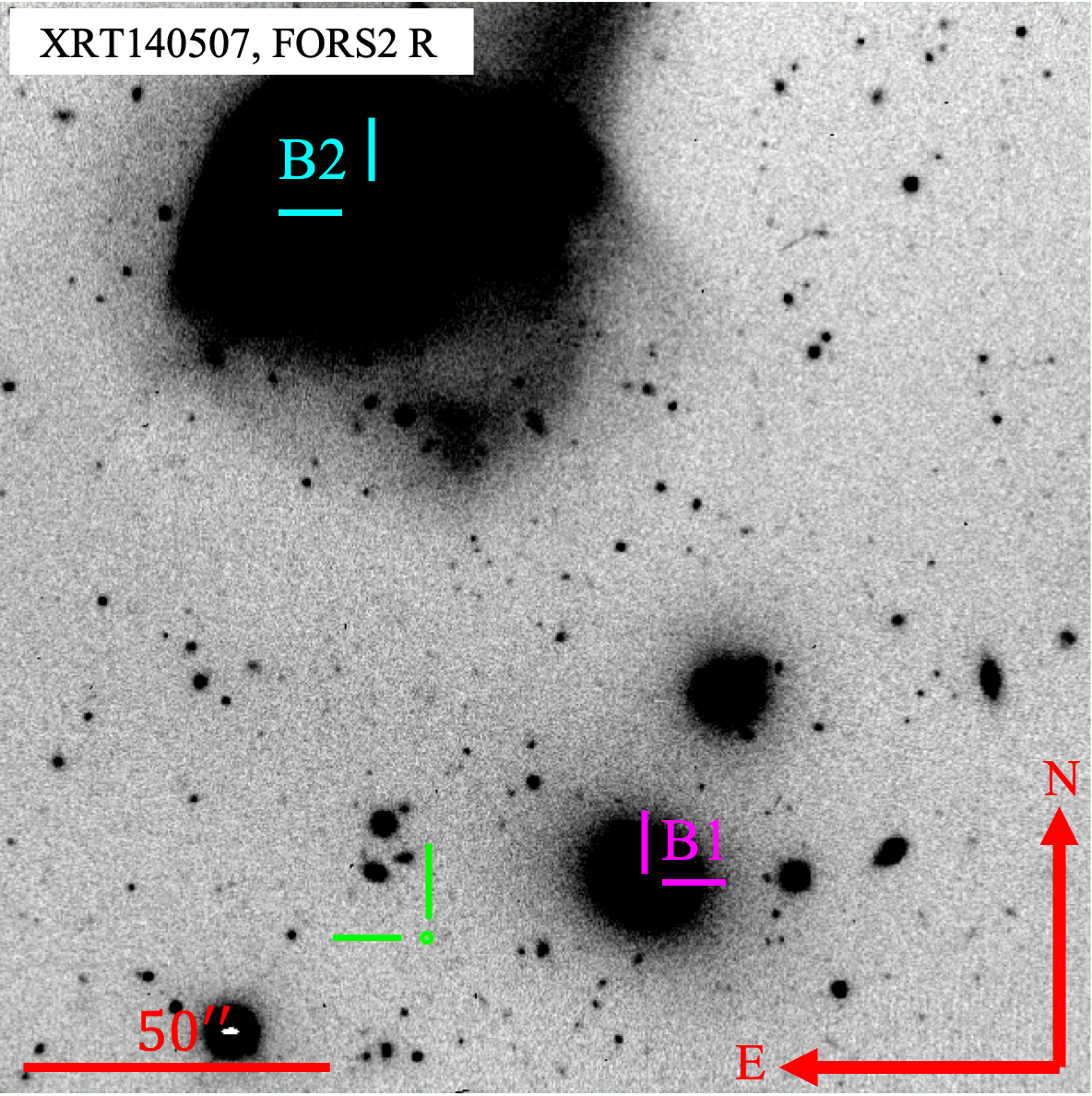}
\end{minipage}
\begin{minipage}[]{0.24\textwidth}
  \includegraphics[width=\linewidth]{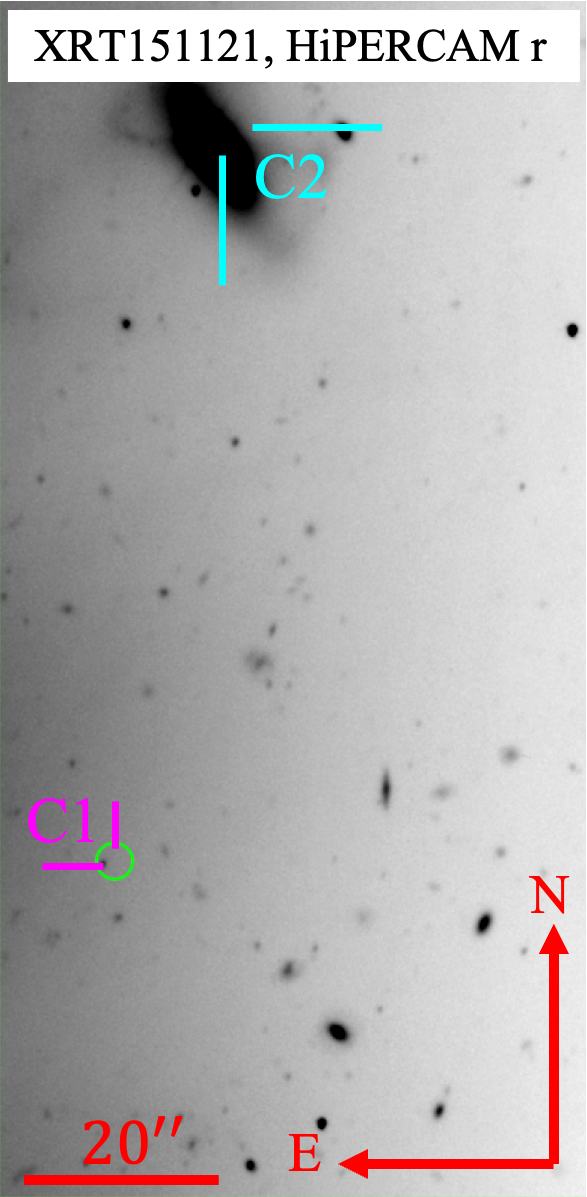}\vspace{0.05cm}
  \includegraphics[width=\linewidth]{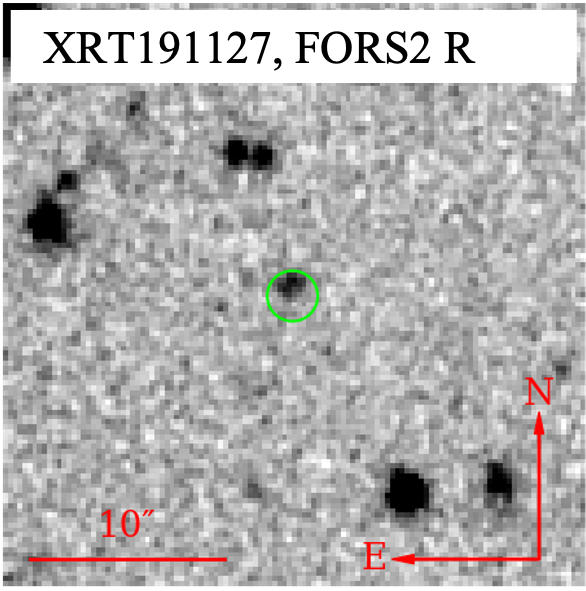}
\end{minipage}
\begin{minipage}[]{0.24\textwidth}
\includegraphics[width=\linewidth]{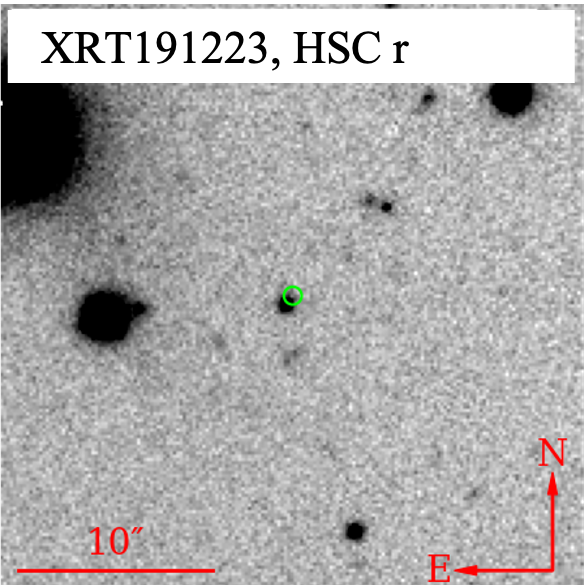}\vspace{0.1cm}
  \includegraphics[width=\linewidth]{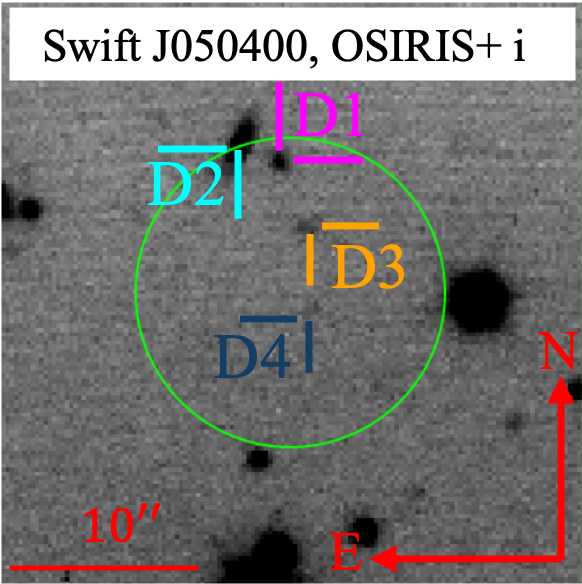}\vspace{0.1cm}
  \includegraphics[width=\linewidth]{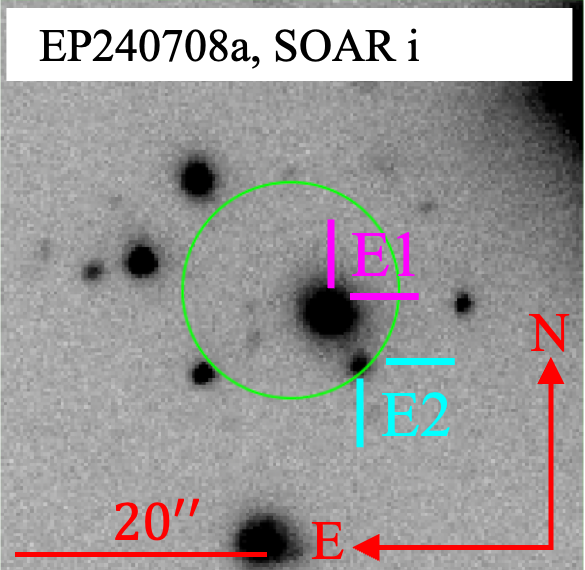}
\end{minipage}
  \caption{Optical images of the FXTs for which no (clear) host galaxy candidate has been reported in the literature before. The names of the FXTs and the instruments and filters are shown at the top of the images. The green circles show the 2$\sigma$ uncertainty regions of the X-ray localisation. The different candidate host galaxies are marked with labels if there is more than one.
    }
    \label{fig:finders}
\end{figure*}

\begin{table*}[ht]
{\small
\centering
\caption{Candidate host galaxy names, coordinates and photometry for all FXTs where a new candidate host was found.}
    \label{table:hostsall}
    \centering
    \begin{tabular*}{\linewidth}{@{\extracolsep{\fill}}c c c c c |c c c c c}
        \hline\hline
Name & RA & Dec & Filter & AB mag & Name & RA & Dec & Filter & AB mag \\ \hline
\multirow{6}{*}{XRT040610 A1} & \multirow{6}{*}{169.5355} & \multirow{6}{*}{7.7047} & g & 22.24 $\pm$ 0.03 & \multirow{7}{*}{XRT151121 C2} & \multirow{7}{*}{40.8265} & \multirow{7}{*}{32.3451} & u & 19.06 $\pm$ 0.05 \\
 &  &  & r & 22.11 $\pm$ 0.01 &  &  &  & g & 17.45 $\pm$ 0.02 \\
 &  &  & i & 21.53 $\pm$ 0.01 &  &  &  & r & 16.49 $\pm$ 0.02 \\
 &  &  & z & 21.31 $\pm$ 0.01 &  &  &  & i & 16.18 $\pm$ 0.02 \\
 &  &  & W1 & 20.89 $\pm$ 0.13 &  &  &  & z & 15.47 $\pm$ 0.03 \\
 &  &  & W2 & 21.55 $\pm$ 0.57 &  &  &  & W1 & 15.40 $\pm$ 0.01 \\ \cline{1-5}
\multirow{6}{*}{XRT040610 A2} & \multirow{6}{*}{169.5379} & \multirow{6}{*}{7.7033} & u & $>$21.03 &  &  &  & W2 & 16.00 $\pm$ 0.01 \\ \cline{6-10} 
 &  &  & g & 22.80 $\pm$ 0.03 & \multirow{7}{*}{XRT191127} & \multirow{7}{*}{207.3471} & \multirow{7}{*}{26.5844} & U & $>$ 21.56 \\
 &  &  & r & 22.61 $\pm$ 0.02 &  &  &  & B & 25.20 $\pm$ 0.10 \\
 &  &  & i & 22.47 $\pm$ 0.02 &  &  &  & V & 23.82 $\pm$ 0.06 \\
 &  &  & z & 22.21 $\pm$ 0.02 &  &  &  & R & 23.60 $\pm$ 0.06 \\
 &  &  & W1 & 20.89 $\pm$ 0.13 &  &  &  & I & 22.90 $\pm$ 0.06 \\ \cline{1-5}
\multirow{8}{*}{XRT040610 A3} & \multirow{8}{*}{169.5485} & \multirow{8}{*}{7.70198} & u & 19.90 $\pm$ 0.08 &  &  &  & W1 & 20.91 $\pm$ 0.11 \\
 &  &  & g & 18.87 $\pm$ 0.03 &  &  &  & W2 & 20.97 $\pm$ 0.28 \\ \cline{6-10} 
 &  &  & r & 18.63 $\pm$ 0.01 & \multirow{2}{*}{Swift J050400 D1} & \multirow{2}{*}{76.0037} & \multirow{2}{*}{67.5702} & g & 23.7 $\pm$ 0.1 \\
 &  &  & i & 18.45 $\pm$ 0.01 &  &  &  & i & 22.8 $\pm$ 0.1 \\
 &  &  & z & 18.34 $\pm$ 0.01 & \multirow{2}{*}{Swift J050400 D2} & \multirow{2}{*}{76.0019} & \multirow{2}{*}{67.5704} & g & 22.7 $\pm$ 0.02 \\
 &  &  & W1 & 18.19 $\pm$ 0.02 &  &  &  & i & 21.2 $\pm$ 0.02 \\
 &  &  & W2 & 18.79 $\pm$ 0.05 & \multirow{2}{*}{Swift J050400 D3} & \multirow{2}{*}{76.0007} & \multirow{2}{*}{67.5690} & g & 24.1 $\pm$ 0.1 \\
 &  &  & W3 & 16.39 $\pm$ 0.18 &  &  &  & i & 23.4 $\pm$ 0.1 \\ \cline{1-5}
XRT140507 B1 & 233.7238 & 23.4713 &  &  & \multirow{2}{*}{Swift J050400 D4} & \multirow{2}{*}{76.0005} & \multirow{2}{*}{67.5677} & g & $>$24.4 \\
XRT140507 B2 & 233.7383 & 23.5037 &  &  &  &  &  & i & 25.2 $\pm$ 0.3 \\ \hline
\multirow{5}{*}{XRT151121 C1} & \multirow{5}{*}{40.8297} & \multirow{5}{*}{32.3239} & u & 25.83 $\pm$ 0.08 & \multirow{3}{*}{EP240708a E1} & \multirow{3}{*}{345.9633} & \multirow{3}{*}{-22.8436} & W1 & 14.57 $\pm$ 0.01 \\
 &  &  & g & 24.55 $\pm$ 0.03 &  &  &  & W2 & 14.00 $\pm$ 0.02 \\
 &  &  & r & 24.26 $\pm$ 0.04 &  &  &  & W3 & 11.16 $\pm$ 0.14 \\
 &  &  & i & 24.36 $\pm$ 0.05 & EP240708a E2 & 345.9625 & -22.8448 &  &  \\
 &  &  & z & 24.65 $\pm$ 0.10 &  &  &  &  &  \\ \hline
\end{tabular*} }
\tablefoot{The J2000 coordinates in RA and Dec are given in column two and three, respectively and their uncertainties are $<<0.1$ arcsec. The optical photometry comes from our work and we obtained W1, W2, and W3 from unWISE where available. Values with `$>$' are 3$\sigma$ upper limits.}
\end{table*}

\begin{figure}[ht]
    \centering
    \includegraphics[width=\linewidth]{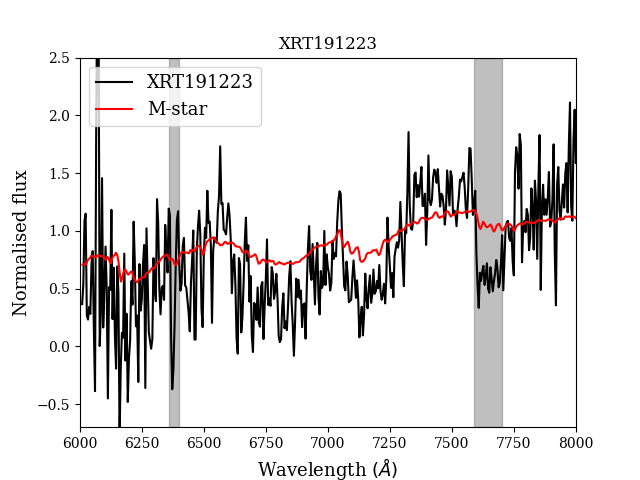}
    \caption{GTC/OSIRIS spectrum of the candidate counterpart of XRT191223 in the wavelength range 6000 to 8000~\AA\ in black with the spectrum of an M1V star over plotted in red. The spectrum of the M1~V star was normalized to its flux at 5556~\AA\ and the spectrum of XRT191223 by the mean flux in the wavelength range shown. We manually masked wavelength region where a strong telluric feature is present (gray shaded area). The M1~V spectral template is taken from \cite{Pickles1998}.}
    \label{fig:xrt191223star}
\end{figure}

-~XRT191223. A faint source was marginally detected before by \cite{Quirola_Vasquez_2023}, we analyse our deeper images and the OSIRIS spectrum to constrain the nature of this source. 
In the deep Subaru HSC \textit{g,r,i,z} images (Fig.~\ref{fig:finders}) the host candidate is consistent with a point source. Additionally we plot the GTC/OSIRIS spectrum of the host candidate together with that of an M1~V-star template spectrum \citep{Pickles1998} in Fig.~\ref{fig:xrt191223star}, and find them to be consistent. Hence, we suggest that XRT191223 is caused by a stellar flare and not by an extra-galactic FXT.
For M-star flares, we expect log$(L_x/L_{bol}) < 0$, although sometimes a so-called "super flare" is seen \citep{Pye2015}. So, as an additional test, we re-calculate the ratio between the X-ray luminosity and the bolometric luminosity (log$(L_x/L_{bol}$)) as in \cite{Quirola_Vasquez_2023}. 
The log$(L_x/L_{bol}$) is calculated by normalizing stellar synthetic models of dwarf stars from \cite{Phillips_2020} to the photometric detections in the \textit{i}-band, and integrating the normalized models over optical wavelengths as a proxy of the bolometric flux. Using the X-ray flare flux of $F_x=1.52 \times 10^{-12}$ \flx, we obtained log$(F_X/F_{bol}) = 0.28 \pm 0.21$. Hence, within a $2\sigma$ error log$(L_x/L_{bol}) < 0$, it is fully consistent with a stellar flare. 
As a final test, we compare its peak X-ray luminosity if it were an M- or brown-dwarf \citep[(L$_{X,peak}^{M-dwarf}\approx$(3.0-300)$\times$10$^{31}$ and L$_{X,peak}^{B-dwarf}\approx$(3.1-300)$\times$10$^{29}$\lum][]{Quirola_Vasquez_2023} and its duration (log($\tau$)$\approx$3.6) with figure 4 in \cite{Mao2025}. We find that it is consistent with the stellar flares in their sample from \cite{Pye2015}. We therefore exclude it from further consideration.

-~Swift~J050400. We detect four objects within the 2$\sigma$ uncertainty region.
For convenience we name the brightest two D1 
and D2, shown with the magenta and cyan markers in Fig.~\ref{fig:finders}, respectively. 
For D1 we find an offset of 8$\pm$3" (16$\pm$8~kpc or 50$\pm$20~kpc, see Section~\ref{subsect:redshift}), and for D2 we find an offset of 9$\pm$3" (50$\pm$20~kpc) between the central position of each galaxy and the centre of the FXT localisation region. The uncertainty arises from the uncertainty in the FXT position.
Additionally, we identify two very faint candidate hosts in the $i^\prime$ band image within the 2$\sigma$ localisation region, marked with orange and navy markers in Fig. \ref{fig:finders} and hereafter called D3 and D4, respectively.  
As we do not have $r-$band detections of these galaxies, we cannot use the formula in \cite{Bloom2002} for the chance alignment, and therefore consider the number of galaxies in a square region of 70\arcsec$\times$70\arcsec in the $i-$band image and assume Poisson statistics to calculate the chance that a source as bright as, or brighter than, the galaxy under consideration within the region is found. We find $P_{chance}=4,9,17$ and 21 percent for D1, D2, D3 and D4, respectively. D1 has the lowest $P_{chance}$ and is therefore most likely to be the true host of Swift~J050400. 

-~XRT140507. No galaxies were detected within the 3$\sigma$ localization uncertainty region in the FORS2 and HiPERCAM images. However, we note that there is a bright galaxy cluster near the position of this FXT, XCLASS~681. The chance alignment probability for the closest galaxy of this cluster, SDSS~J153453.70+232816.5, with an offset of 39$\pm$1\arcsec ($\sim$70~kpc) is 3.3\%.
It has a spectroscopic redshift of 0.09025~$\pm$~0.00002 \citep{Ahn2012}. For convenience, this galaxy is referred to as B1 hereafter.
Additionally, SDSS~J153457.20+233013.2 is considered as a potential host galaxy, with an offset of 127$\pm$1\arcsec ($\sim$50~kpc) and  chance alignment probability of $\sim$4.3\%. This galaxy has a catalogued spectroscopic redshift of 0.01838$\pm$0.00002. Hereafter, we call this galaxy B2.

-~XRT040610. We identify three candidate host galaxies, all located outside the 2$\sigma$ uncertainty region of the FXT. The galaxy, hereafter called A1, with the lowest chance alignment probability, although still relatively high, of 14\% is at 8$\pm$1\arcsec ($\sim$65~kpc) offset. This galaxy has a photometric redshift in SDSS-DR18 \citep{Almeida_2023}, 
but since the photometry for this galaxy is flagged as likely unreliable, we do not adopt this redshift in this work.
Followed by A2 with P$_{chance} \approx$ 15\%, it has an offset of 6$\pm$3~\arcsec (50$\pm$25~kpc). The last considered galaxy, A3, is SDSS~J111811.64+074207.1 with an offset of 44$\pm$1\arcsec ($\sim$70~kpc) and chance alignment probability of 25\%. A3 has a photometric redshift of 0.08$\pm$0.02 \citep{Almeida_2023}.

-~XRT151121. One candidate host galaxy is identified within the 2$\sigma$ uncertainty region, hereafter referred to as C1 with a chance alignment probability of $\sim$5\%. We also note the presence of a bright ($m_r \approx 15.8$ in Pan-STARRS) galaxy at 76$\pm$2\arcsec ($\sim$~60~kpc) offset of the FXT location, hereafter called C2. C2 has a photometric redshift of 0.04$\pm$0.01 \citep{Almeida_2023}.
As the chance alignment probability is $\sim$15\% for C2, we consider C1 as the most likely host galaxy of this FXT.

-~EP240708a. We identify two galaxies within the 3$\sigma$ uncertainty region of EP-FXT in the SOAR i-band images, Fig.~\ref{fig:finders}. A bright galaxy, with $m_r$~$\sim$~19 AB mag, and a fainter one, $m_r$~$\sim$~21 AB mag are identified in Pan-STARRS and in our images. We refer to them as E1 and E2, respectively. The chance alignment probabilities are 3\% and 14\% for E1 and E2, respectively. .

\subsection{Photometry}
\label{sect:phot}
We do (forced) photometry on the candidate host galaxies of Swift~J050400, XRT191127, XRT040610 and XRT151121, using SE. We adjust the seeing, pixel scale, saturation level and gain values following the SE manual\footnote{\url{https://sextractor.readthedocs.io/en/latest/index.html}} and keep the values of the remainder of the SE parameters at the default. For the galaxies that are observed in different filters by the same telescope/instrument, we use the filter-image with the deepest limiting apparent magnitude as the detection image. Then, on the other filters the photometry is done in dual-image mode, where the source is detected in the detection image and the measurements are done on the second image. The resulting Kron magnitudes \citep[][]{Koo_1986, Kron_1980} (returned by SE as \texttt{MAG\_AUTO}) are taken as the galaxies' photometry, and given in Table~\ref{table:hostsall}. For the \textit{griz}-filters, we use Pan-STARRS to calibrate the photometry while for the \textit{u}-band we use SDSS. The FORS2 filters are calibrated against SDSS stars in the field and using the \cite{Jordi_2006} transformations to transform these to the UBVRcIc system. Upper limits (3$\sigma$) are determined using the python package Photometry Sans Frustration \citep{Nicholl_2023}, we note that they show the upper limit on a point source and not on an extended source. 
Additionally, for XRT191127, for all candidate host galaxies for XRT040610, XRT151121 C2 and EP240708a E1 we use the available unWISE magnitudes \citep{Lang_2014, Lang_2016}. They are also given in Table~\ref{table:hostsall}.

\subsection{Galaxy fitting}
\label{Section:fitting}
\begin{table*}[ht!]
\caption{Properties of the candidate host galaxies obtained from fitting the photometric SED and/or spectrum. 
}
\label{table:fittingparams}      
\centering 
\begin{tabular*}{\linewidth}{@{\extracolsep{\fill}}l c c c c c c c c}   
\hline\hline 
 Galaxy & $z$ & L$_{X,peak}$& M$_*$ & SFR & Z & Age & A$_V$& \\ 
  &  & ($\times 10^{44}$ \lum)& ($\times 10^{10}$ M$_{\odot}$) &  (M$_{\odot}$ yr$^{-1}$) & (Z$_{\odot}$) & (Gyr) & & \\ 
 \hline 
XRT040610 A1                & 0.9(1)$^{p}$                 & $2.02\pm0.38$                        & 1.4$\pm$0.2        & 9$\pm$2         & 0.4$\pm$0.1          & 1.2$\pm$0.4               & $0.1\pm0.1$    &  \\
XRT040610 A2                & 1.6(1)$^{p}$                 & $7.9 \pm 1.4$    
& 2.8$^{+1.3}_{-0.7}$          & $80 ^{+10}_{-20}$       & 1$^{+1}_{-0.5}$   & 0.3$\pm$0.1               & 0.8$\pm$0.1               &  \\
XRT040610 A3                & 0.05(2)$^{p}$    & 0.0028(5)    & 0.036(3)   & 0.6$\pm$0.1  & 1.1$\pm$0.4 & 0.46(6)               & 0.7$\pm$0.1            &  \\
XRT100424                   & 0.1820(2)$^{s}$              & 0.054(11)                              & 12.5$^{+2.2}_{-1.2}$           & $<1.4$                  & 1.2$^{+0.5}_{-0.4}$ & 2.58$^{+0.97}_{-0.35}$ & 0.07$^{+0.07}_{-0.05}$ &  \\
XRT140507 B1                & 0.09025(2)$^{s,\dagger}$ & 0.041(20)                              & 5.8$^{+1.0}_{-0.2}$         & 0.12$\pm$0.01           & 2.22$^{+0.07}_{-0.20}$ & 5.32$^{+1.77}_{-0.27}$ & 0.31(2)                &  \\
XRT140507 B2                & $0.01839(1)^{s,\dagger}$ & 0.0016(74)                             & 0.74$\pm$0.01                  & 0.32$\pm$0.01           & 2.42$\pm$0.02          & 8.76$^{+0.25}_{-0.06}$ & 2.00(1)                &  \\
XRT151121 C1                & 2.87(3)$^{p}$                 & $900 \pm 350$                          & 0.95$^{+0.60}_{-0.40}$         & 12$\pm$2              & 1.7$\pm$0.5        & 0.6$\pm$0.3 & 0.1$\pm$0.1 &  \\
XRT151121 C2    & 0.04(1)$^{p}$    & 0.045(25)   & 1.7$\pm$0.4   & 0.5$\pm$0.1     & 0.8$\pm$0.2 & 7.8$^{+0.9}_{-1.5}$    & 1.1$\pm$0.1    &  \\
XRT191127                   & 0.32$^{+0.03}_{-0.07}$$^{p}$ & 50$\pm$25                       & 0.3$\pm$0.1                    & 1.1$^{+0.3}_{-0.5}$  & 2.2$^{+0.1}_{-0.4}$ & 1.8$^{+1.0}_{-0.6}$ & 1.9$\pm$0.1 &  \\
Swift J050400 D1            & 0.106(1)$^{s}$               & 0.12$\pm$0.07                          & 7.5$\pm$0.1                    & $<0.03$                 & 0.07$^{+0.13}_{-0.04}$ & 5.5$\pm$2.2            & 0.13$^{+0.14}_{-0.09}$ &  \\
Swift J050400 D1            & 0.450(2)$^{s}$               & 3.2$\pm$1.9                            & 8.7$\pm$0.1                    & 0.3$^{+0.2}_{-0.1}$  & 2.3$^{+0.1}_{-0.2}$ & 1.4$^{+0.9}_{-0.6}$    & 0.3$^{+0.3}_{-0.2}$ &  \\
Swift J050400 D1            & 0.494(1)$^{s}$               & 4.0$\pm$2.4                            & 8.8$\pm$0.1                    & 0.3$\pm$0.1  & 2.2$^{+0.2}_{-0.8}$ & 1.7$^{+1.0}_{-0.7}$ & 0.21$^{+0.25}_{-0.15}$ &  \\
Swift J050400 D1            & 0.950(1)$^{s}$              & 20.0$\pm$11.9                          & 8.7$\pm$0.1                    & 2.7$^{+1.0}_{-0.6}$     & 2.3$\pm$0.1            & 0.11(4)                & 0.23$^{+0.18}_{-0.13}$ &  \\
Swift J050400 D2            & 0.3663(1)$^{s}$              & 2.0$^{+3.4}_{-1.2}$                    & 0.20$\pm0.03$                  & 0.12$\pm$0.04  & 0.13$\pm$0.02          & 3.58$^{+1.98}_{-1.20}$ & 0.66(19)               &  \\
EP240708a E1   & 0.3734(2)$^{s}$              & $\sim$5400                             & 1.07$\pm$0.02  & 0.24(1)                 & 0.10$\pm$0.02  & 6.9$\pm$0.2                  & 0.02$\pm$0.01                &  \\
EP240708a E2  & 0.3728(1)$^{s}$   & $\sim$5400        & 8.74(1)      & 0.045(2)    & 2.22$\pm$0.06      & 1.43(3)                & 0.06(3)   &  \\ \hline                  
\end{tabular*}
\tablefoot{The first column lists the FXT and the candidate host galaxy name. The second column lists the redshift ($z$), where the digit in brackets indicates the uncertainty in the last digit and the marker indiciated whether it is a spectroscopic $(s)$ or photometric $(p)$ redshift. The redshift values marked with a $\dagger$ are taken from the literature, see references in the text. In the third column we provide the peak X-ray luminosity of the FXT assuming the candidate host on this row is the real host. The fourth column shows the galaxy's stellar mass (M$_*$), the fifth column its star formation rate (SFR), the sixth column the metallicity (Z) in terms of solar metallicity, the seventh column the mass-weighted age of the galaxy, and the last column the dust attenuation parameter. }

\end{table*}

We fit the spectra of the candidate host galaxies for XRT100424, Swift~J050400~D1 and D2, and EP240708a and the photometric SED of XRT191127, XRT040610 and XRT151121 using the Python fitting package: Bayesian Analysis of Galaxies for Physical Inference and Parameter EStimation \citep[BAGPIPES,][]{bagpipes}. For both candidate host galaxies of XRT140507 we fit the spectra obtained through SDSS-DR18 \citep{Almeida_2023}. An exponentially declining star formation history \citep{Simha_2014,Carnall_2019} is considered and we fit for the exponential timescale $\tau$, the formed stellar mass, age, redshift (z), metallicity (Z), and dust extinction ($A_V$) as from \cite{Calzetti2000}.
The best fits are shown in Appendix~\ref{Section:AppendixSEDfit}. The data are shown in blue with the best fit over plotted in orange. The posterior distribution of five of the fitted parameters is shown in the bottom panels.
Due to the low signal-to-noise for the spectrum of Swift~J050400~D1 and since there is only one line visible in the spectrum, we try different redshifts interpreting this line as H$\alpha$ ($z=0.11$), [OIII] ($z=0.45$), H$\beta$ ($z=0.49$), or [OII] ($z=0.95$).
We also note that the fits of the spectrum of the host galaxy EP240708a E1 seems to underestimate the peak of the H$\alpha$ and [SII] emission lines. This might be explained by AGN activity in this galaxy. We use the $W1 - W2$ colour to test if it is consistent with an AGN using the selection criteria of \cite{Stern2012, Assef2013, Assef2018}. We find $W1 - W2 = 0.57 \not> \alpha_R \mathrm{exp}\{\beta_R(W2 - \gamma_R)^2\}=0.66$ for $\alpha_R=0.662$, $\beta_r=0.232$, and $\gamma_R=13.97$ and is therefore not consistent with an AGN at 90\% confidence.

The obtained best-fit parameters are given in Table~\ref{table:fittingparams}. We use the 16th, 50th and 84th quantiles of the posteriors from the fit for the redshift, galactic stellar mass, star formation rate (SFR), metallicity, mass-weighted age (Age$_{MW}$) and A$_V$. For both candidates of XRT140507 a spectroscopic redshift is known and hence, we do not fit for the redshift of those galaxies but instead we keep the redshift fixed at their predetermined values. For XRT040610~A3 and XRT151121~C2 a photometric redshift is known ($z_{phot}=0.08\pm0.02$ and $z_{phot}=0.04\pm0.01$, respectively) and we find similar results for the other parameters when fitting both with the redshift fixed to these values and when leaving it as a free parameter. The values in the Table are obtained by leaving the redshift as a free parameter, and our best-fit redshift is consistent with the photometric redshift in each case.

\section{Discussion}
\label{SectDiscussion}
We obtained deep optical images to find and study the properties of the host galaxies of the FXTs Swift~J050400, XRT191223, XRT140507, XRT040610, XRT151121, and EP240708a. Furthermore, we obtained spectra of the host candidates of Swift~J050400, XRT100424, XRT191223, and EP240708a to further study those candidate host galaxies. 
Our findings show that the event XRT191223 is consistent with originating from a stellar flare, therefore, we exclude it from further investigations. 

\subsection{Redshifts and energetics}
\label{subsect:redshift}
We report spectroscopic or photometric redshifts for the candidate host galaxies of the sample of FXTs in Table~\ref{table:fittingparams}. 
We use these redshifts to calculate the peak X-ray luminosities ($L_{X,peak}$) from the measured FXT peak X-ray fluxes, see Table~\ref{table:fittingparams} \citep{AlpLarsson2020, Quirola_Vasquez_2022, Quirola_Vasquez_2023, Evans2023, 2024GCNEP240708a}, under the assumption that the candidate host is indeed the actual host galaxy.
We then compare the preliminary peak X-ray luminosities of the FXTs with the various proposed progenitor models. The progenitor models considered are (cocoon emission from) LGRBs \citep[$L_{X,peak}\gtrsim 10^{47}$~erg~s$^{-1}$,][]{Gottlieb_2022}, millisecond magnetar spin down luminosity SGRBs/BNS mergers \citep[$L_{X,peak} \sim 10^{44} - 10^{47}$~erg~s$^{-1}$,][]{Berger_2014}, SN SBOs \citep[$L_{X,peak}\sim 10^{42} - 10^{44}$~erg~s$^{-1}$,][]{Waxman_2017}, and WD–IMBH TDEs \citep[$L_{X,peak}^{non-relativistic}\lesssim 10^{43}$~erg~s$^{-1}$, $L_{X,peak}^{relativistic} \sim 10^{43}-10^{48}$~erg~s$^{-1}$,][]{Maguire_2020}, although we note that the expected rates for the latter are low \citep{Maguire_2020}.
The potential peak X-ray luminosities of our sample cover a large range, with XRT100424, Swift~J050400~D1 (if not at $z=0.95$), XRT140507, XRT040610~A3 and XRT151121~C2 being on the low end and therefore being consistent with SN SBOs, non-relativistic WD-TDEs and perhaps overlapping with the low end of the peak energies of BNS mergers. 
Notably, these peak luminosities would be in the order of $\sim10^{42}$\lum~ similar to the peak luminosities of the transient class Hyper-Luminous X-ray sources \citep[HLXs, $L_{X,peak}\ge10^{41}$\lum; e.g.][]{Farrell2009, Heida2015}. These authors show that HLXs are strong candidates for IMBHs since their luminosities can be explained by accretion onto black holes with masses of $M>100M_{\odot}$ and their X-ray spectra show evidence for relatively low blackbody temperatures, corresponding to BH masses $M = 10^2 - 10^5M_{\odot}$. HLXs show long-term variability. 
For the FXTs in our sample with similar peak luminosities, it cannot be ruled out that they are variable over long time periods, due to the low number of observations \citep{AlpLarsson2020, Quirola_Vasquez_2022, Quirola_Vasquez_2023}, and therefore they can be candidates for HLXs.
The other potential peak X-ray luminosities are more consistent with a BNS merger or relativistic WD-TDE origin. Only the peak luminosities of XRT151121~C1 and EP240708a come close to the peak luminosity of LGRBs, with L$_{x,peak}\sim10^{47}$~\lum.

\subsection{Offsets between FXT and host galaxy candidate}
\label{sect:offset}
\begin{figure}[ht!]
    \centering
\includegraphics[width=\linewidth]{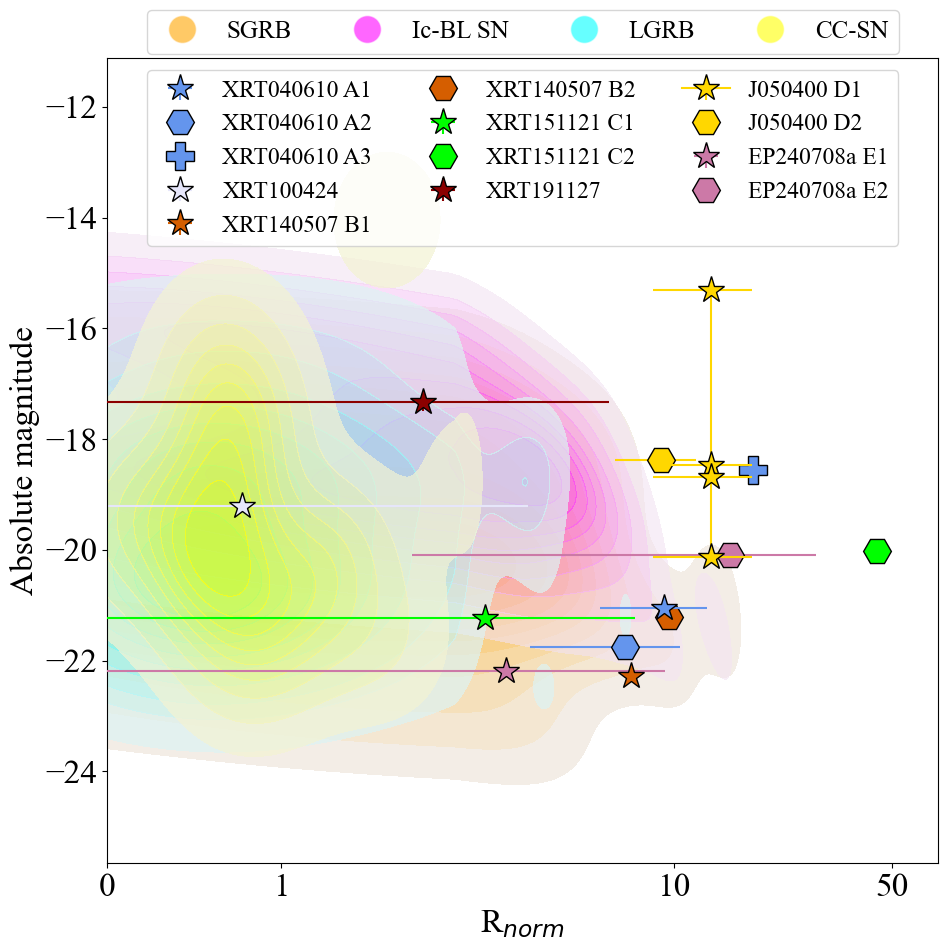}
    \caption{Host normalised offset $R_{norm}$ (in symmetric logarithmic scale) versus host absolute $r^\prime$ magnitude  for our sample of FXTs compared to different transient classes: SGRBs \citep{Fong_2022}, Ic-BL SN \citep{Japelj2018}, LGRBs \citep{Bloom2002, Blanchard2016}, and CC-SN \citep{Kelly_2012, Schulze_2021}, shown as green, magenta, cyan, and yellow contours, respectively. The candidate host galaxies for the FXTs in our sample are shown with coloured markers, where we assigned a star for the galaxy we identified with the lowest $P_{chance}$, if we identified more than one candidate, the second $P_{chance}$ value is shown with a hexagon, and the third with a plus-sign. Each FXT is shown with a different colour. All four redshift possibilities for Swift~J050400~D1 are shown on one line.
    Depending on which candidate host is the actual host, all FXTs share overlap within 1 or 2$\sigma$ with LGRBs, SGRBs and CC-SNe, whereas only XRT191127 and XRT100424 also overlap with Ic-BL SNe. Note that for some or even all FXTs under discussion here, it is possible that the true host galaxy has not yet been detected.
    }
    \label{fig:offset}
\end{figure}

We calculate for each FXT the physical offset between the center of the candidate host galaxy/galaxies and the center of the FXT localization uncertainty region. In Fig.~\ref{fig:offset} we plot the host normalised offsets $R_{norm}$, i.e.~the offset divided by the half-light radius ($R_{h}$) of the host galaxy, versus the absolute $r^\prime$-band host magnitude. We compare with the normalised offset for the host galaxies of SGRBs \citep{Fong_2022}, LGRBs \citep{Bloom2002, Blanchard2016}, Type Ic broad-lined (BL) SN \citep{Japelj2018} and core collapse SN \citep{Kelly_2012, Schulze_2021}. 
In order to calculate the absolute $r-$band magnitudes we use the apparent magnitudes in Table~\ref{table:hostsall} and in addition for the host galaxies of XRT100424, XRT140507, XRT151121~C2 and EP240708a, we use the apparent magnitudes from Pan-STARRS1 and from the Legacy Survey for XRT191127, combined with the redshifts from Table~\ref{table:fittingparams}. For the two candidate host galaxies of Swift~J050400, we only have $g^\prime$ and $i^\prime$ apparent magnitudes from our photometric observations. Here, we measured the apparent $r^\prime$-band magnitudes using the GTC/OSIRIS+ spectra employing the python package \texttt{PYPHOT} \citep{zenodopyphot}.
The half-light radii are taken from the Legacy Survey data release 10, except for Swift~J050400 and XRT151121 which are not covered by the Legacy Survey. 
For the candidate host galaxies of XRT151121 the half-light radii are found using the \texttt{FLUX\_RADIUS} output from SE on the $r$-band image and for Swift~J050400 on the $i$-band image, as that is the image closest to the $r$-band. 
The uncertainties on $R_{norm}$ arise from the uncertainties in the FXT localisations and on the measurements of the half-light radii. 
Distribution functions for the physical offsets of CC-SNe show a maximum projected distance of 37~kpc \citep{Schulze_2021}. Whereas BNS systems can obtain high systemic velocities from two potential velocity kicks received when the neutron stars form in the SN explosions \citep{Andrews_2019}. There have been observations of BNS mergers with offsets as large as $\sim$~60~kpc \citep[e.g.,][]{Fong_2022}. As IMBHs are thought to reside in globular clusters or (ultra-compact) dwarfs, the offsets of IMBH-WD TDEs can span a wide range, depending on whether the dwarf host galaxy is detected or rather the main galaxy that might have a companion dwarf galaxy.

In Fig.~\ref{fig:offset} we see that XRT100424, XRT191127, XRT151121 C1 and EP240708a are consistent within their uncertainties with all transients classes we compare to. We also note that for the galaxies with the largest $R_{norm}$, we identified another candidate host with a lower $P_{chance}$, except for Swift~J050400. We also note that for all candidate host galaxies for each FXT, except for Swift~J050400, the galaxy with the largest $R_{norm}$, also has the largest offset.

The larger $R_{norm}$ of Swift~J050400, XRT040610 and XRT140507 might be explained by the way the FXTs were selected, namely not having a (clear) host candidate in their uncertainty region before this work, which indicates that the hosts are either faint or far offset. 
To test whether the FXT candidate host offsets are drawn from the same 2D $R_{norm}$--absolute magnitude distribution as any of the transient classes we compare with, we perform a 2D Kolmogorov-Smirnov (KS) test on the candidate host galaxies with the lowest $P_{chance}$ for each FXT. We find $p=0.2$ for the SGRB offset distribution. Hence, we cannot discard the hypothesis that the samples are taken from the same underlying distribution. For all other transient classes $p<<0.05$, allowing us to discard the hypothesis that these samples are drawn from one underlying distribution.

\subsection{SFR versus M*}
\begin{figure}[ht]
\centering
\includegraphics[width=\linewidth]{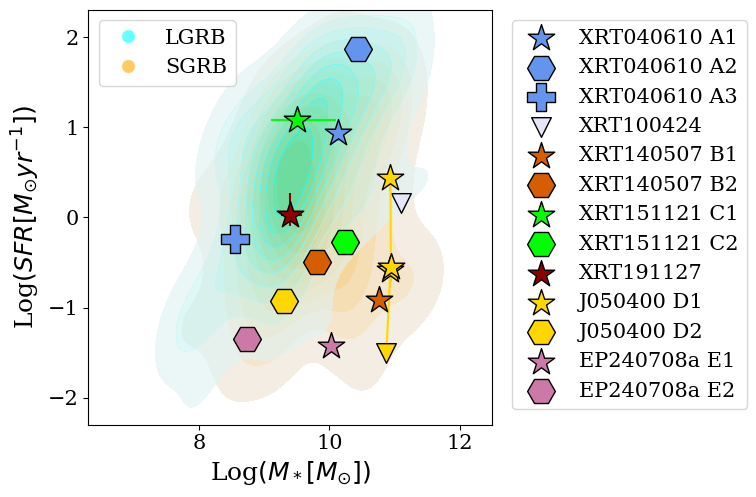}
  \caption{Total stellar mass (M$_*$) and star formation rate (SFR) of the candidate host galaxies of our sample of FXTs compared to those of LGRBs and SGRBs, shown with cyan and orange contours, respectively. The coloured markers for the FXT in our sample are the same as in Fig.~\ref{fig:offset}. For Swift~J050400~D1, we show all possibilities for all possible redshifts on one line. Most FXTs are consistent with the host galaxies of both LGRBs and SGRBs, except for XRT100424, XRT140507~B1, Swift~J050400~D1 and EP240708a~E1 which are only consistent with the SGRB hosts.}
     \label{fig:SFRvsM}
\end{figure}

In Fig.~\ref{fig:SFRvsM} we plot the total stellar masses of the galaxies (M$_*$) versus the SFR values inferred with BAGPIPES of the candidate host galaxies of the FXTs compared to the host galaxies of LGRBs \citep{chang_2015, Li_2016, Izzo_2020, Ho_2020} and SGRBs \citep{Li_2016, Im_2017, Nugent_2022}. For Swift~J050400~D1 we plot all four possibilities given the potential redshifts (as described in Section~\ref{Section:fitting}). The host galaxies of XRT100424, Swift~J050400~D1, XRT140507~B1 and EP240708a~E1 share no overlap with LGRB host galaxies, as they have less star formation, or higher masses, but can be consistent with BNS merger host galaxies. 

IMBH-WD TDE host galaxies are not in the Figure, as there are no observational data of this transient class. However, their SFR is expected to be very low if the IMBH is in a globular cluster, while dwarf galaxies on the other hand can have very high SFRs. If the mass of the galaxy is significantly higher than $\sim10^{9}$~$M_{\odot}$, it is unlikely to harbour an IMBH at the center, as the central black hole mass would be higher than $10^{5}$~$M_{\odot}$ \citep{McConnell2013}. From Fig.~\ref{fig:offset}, we see that for all FXTs with M$_*>$10$^{9}$, the offset is large enough \citep[$R_{norm}\gtrsim1$, e.g.][]{Bhandari2022} that an unseen globular cluster harbouring the IMBH associated with this main galaxy is a feasible explanation.

\subsection{Underlying (dwarf) galaxy limits}
\label{Sec:undetectedgalaxies}
For the FXTs without a candidate host inside the 1$\sigma$ localisation, we calculate the magnitude limits for a possible galaxy in this region. From this, we discuss the likeliness that the true host is undetected. 
To assess the possibility of a faint host at the FXT position, we perform various stacking methods using the images obtained in different filters for the fields of XRT140507, XRT040610 and Swift~J050400, including creating a deep white light image and stacking only blue and only red filters. In the case of XRT040610, one can see a faint possible object in the uncertainty region in the white light image (Fig.~\ref{fig:xrt040610_ugriz}), but there are not significantly more counts than in the background, and thus deeper observations are needed to confirm it.

For both XRT140507 and XRT040610, the deepest 3$\sigma$ limiting magnitudes (as described in Section \ref{sect:phot}) are $g>25$~mag and for Swift~J050400 $g>24.4$~mag. If we assume an absolute $g$-band magnitude for a dwarf galaxy of $-$15~mag, the redshift of an undetected underlying dwarf galaxy would have to be $z>0.22$ for both XRT140507 and XRT040610 and $z>0.17$ for Swift~J050400. The implied lower limits on the peak X-ray luminosities do not rule out any considered origin for either of these FXTs.
For an undetected spiral galaxy of absolute magnitude $-$18~mag, the redshift would have to be $z>0.9$ or $z>0.7$ for XRT140507 and XRT040610 and Swift~J050400, respectively. This would start to rule out the SN SBO and non-relativistic WD–IMBH TDE progenitors, as then $L_{X,peak}\gtrsim10^{44}$~\lum. 

\begin{figure}
    \centering
    \includegraphics[width=\linewidth]{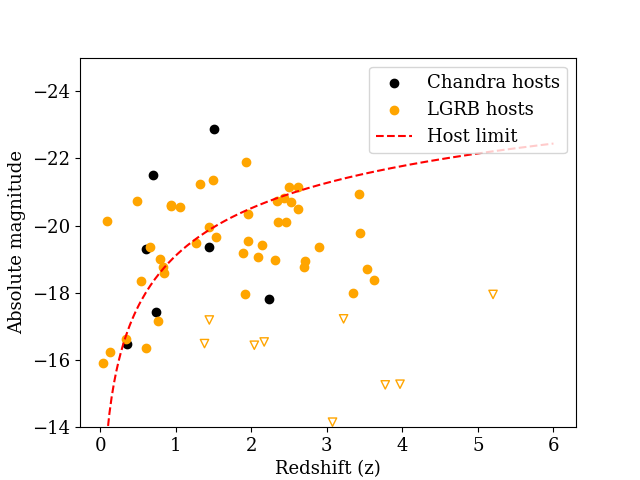}
    \caption{Absolute R band magnitude versus redshift for the sample of LGRB host galaxies from the TOUGH survey \citep{Hjorth2012,Schulze2015}) and \emph{Chandra} FXTs \citep{Quirola_Vasquez_2022,Quirola_Vasquez_2023} in orange and black, respectively. Upper limits are plotted as open triangles. The limiting $r-$band magnitude of our images of the field of XRT140507 and XRT040610 converted to an absolute magnitude given an assumed redshift is indicated by the red dashed line. For $z\gtrsim2.5$ we would not detect typical LGRB host galaxies in our images.}
    \label{fig:absmagvsz}
\end{figure}

We also compare the absolute magnitude (limits) of our observations with that of LGRB host galaxies derived from the TOUGH sample \citep{Hjorth2012, Schulze2015}. We plot the absolute $r-$band magnitude versus redshift in Fig.~\ref{fig:absmagvsz} of this sample, together with the absolute magnitude of the samples in \cite{Quirola_Vasquez_2022,Quirola_Vasquez_2023} and the limiting absolute magnitude derived from our images of $r\gtrsim24$. From this Figure we can see that at $z\gtrsim2.5$ we would not detect a typical LGRB host galaxy. Therefore, for XRT140507, XRT040610 and Swift~J050400 a similar origin as LGRBs cannot be ruled out as we might not be seeing the true host galaxies of these FXTs. In general, if the true host galaxy is shown to be at a significantly larger redshift than $z=2.5$, we cannot rule out several of the potential progenitor models.

We note that Swift~J050400~D3 and Swift~J050400~D4 are marginally detected objects within the uncertainty region of Swift~J050400 and we test if either of them could be a dwarf galaxy, by calculating their (3$\sigma$ upper limits on the) absolute $g$-band magnitudes and half-light radii in parsec for different redshifts to see if they fall on the size-luminosity relation for dwarf galaxies \citep[as e.g.~in][]{Simon2019, Eappachen_2022}. From the resulting plot (shown in Fig.~\ref{fig:Swift_D3D4_dwarfs}) we can see that neither D3 nor D4 falls on the (extrapolated) size-luminosity relation, except for the region between $log(R_h)\sim2.2$ and $log(R_h)\sim2.5$, which corresponds to $0.006 \lesssim z_{D3} \lesssim 0.012$ and $0.01 \lesssim z_{D3} \lesssim 0.02$. However, at these redshifts, the peak X-ray luminosity of Swift~J050400 would be L$_{X,peak}$~$\lesssim 10^{41}$~\lum, which is rather low for all proposed origins. We note that since the magnitudes for D4 are 3$\sigma$ upper limits, there could be a wider range of possible redshifts for this galaxy to be  consistent with a dwarf galaxy.
Additionally, we follow the method described by \citet{Eappachen_2022} to see at what redshifts D3 and D4 would be on the size-luminosity relation from \cite{zhang2019}. We find that the size-luminosity relation for D3 is consistent with the observed spiral galaxy size-luminosity relation for redshifts in the range of 0.7-1.3 or 3.2-3.9. For D4 this range is 1.4-2.3, see Fig.~\ref{fig:Swift_D3D4_spirals}. These redshifts rule out the SN SBO and non-relativistic WD–IMBH TDE progenitor as the peak X-ray luminosity would be higher than expected for those scenarios.

\subsection{Comparison to other FXTs}
\begin{figure}[ht]
    \centering
    \includegraphics[width=\linewidth]{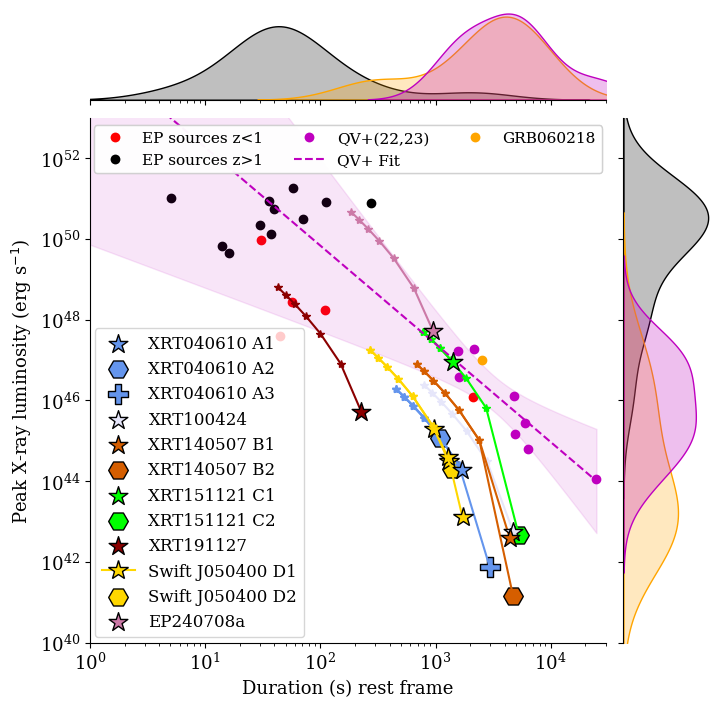}
    \caption{Rest-frame duration of the FXT in seconds versus its peak X-ray luminosity. We compared it to several EP-discovered FXTs (from the sample of \citealt{Guo2025}; black $z>$1 and red $z<1$ circles) and to FXTs detected by \emph{Chandra} (\citealt{Quirola_Vasquez_2022, Quirola_Vasquez_2023}; magenta circles). The magenta line shows a power-law fit through the \citet{Quirola_Vasquez_2022, Quirola_Vasquez_2023}--\emph{Chandra} detected sample with the (extrapolated) 3$\sigma$ uncertainty indicated by the butterfly pink hue. The coloured markers for the FXT in our sample are the same as in Fig.~\ref{fig:offset}. While the FXT duration of our sample is similar to the one of the \citet{Quirola_Vasquez_2022, Quirola_Vasquez_2023}--\emph{Chandra} detected sample, the peak X-ray luminosities are clearly different while there is no data selection or other reason why they should be different. Since the true host galaxy might be a much fainter and higher redshift source, we plot a line for each FXT in our sample with increasing redshift up to $z=6$. At the top and right-hand side of the plot, we show the distributions in duration and peak X-ray luminosity with black, orange and magenta for all the EP-discovered FXTs, the FXTs of this sample and the \citet{Quirola_Vasquez_2022, Quirola_Vasquez_2023}--\emph{Chandra} detected sample, respectively. The p-value of a 2D KS test for the EP and FXT distributions is $p=0.0001$, the p-value for the \emph{Chandra} and this FXT sample is $p=0.13$.}
    \label{fig:lum_duration}
\end{figure}

In Fig.~\ref{fig:lum_duration} we plot the rest-frame duration versus the peak X-ray luminosity for the FXTs of this sample together with EP-discovered FXTs that have a reported peak X-ray flux from \cite{Guo2025} and \emph{Chandra}-discovered FXTs from \cite{Quirola_Vasquez_2022,Quirola_Vasquez_2023} with redshifts from host studies. The two samples cover a different region of the parameter space. We plot the distributions in durations and peak X-ray luminosity for the EP sample, the candidate host galaxies of this sample with the lowest $P_{chance}$ and the \citet{Quirola_Vasquez_2022, Quirola_Vasquez_2023}--\emph{Chandra} detected sample at the sides of Fig.~\ref{fig:lum_duration}. To test if these distributions can be drawn from the same underlying distribution, we perform 2D KS tests. For the distributions of the EP sample and our sample, we find $p=0.001<<0.05$, indicating a very low statistical probability that these are the same. For the \citet{Quirola_Vasquez_2022, Quirola_Vasquez_2023}--\emph{Chandra} detected sample and our sample, we find $p=0.13$, indicating that they could be the same distribution. We discuss the differences in the following subsections.

\subsubsection{Chandra detected FXTs}
From Fig.~\ref{fig:lum_duration}, we note that the \emph{Chandra}-discovered FXTs all have a similar duration of a several hundred to a few thousand seconds. We also note that those discussed in \citet{Quirola_Vasquez_2022, Quirola_Vasquez_2023} tend to be somewhat more luminous than the FXTs in the sample discussed in this paper. However, they are not as luminous as the EP-discovered FXTs. These differences are unexpected and we investigate the possibility that the FXTs from our sample actually originate from galaxies at a higher redshift than the candidate host galaxies we associated with the events. In Fig.~\ref{fig:lum_duration} the peak luminosity and rest-frame duration is shown as a function of increasing redshift for the FXTs in our sample starting from the lowest redshift of the candidate host galaxy, up to $z=6$. We fit a power law to the \emph{Chandra}-discovered FXTs discussed in \citet{Quirola_Vasquez_2022, Quirola_Vasquez_2023} and overplot the best-fit power law with its 3$\sigma$ butterfly uncertainty region and see if the FXTs from our sample cross this region for any redshift. XRT151121 and EP240708a fall right in this uncertainty region. Swift~J050400 and XRT140507 only overlap with the 3$\sigma$ uncertainty region after $z\approx5$ while XRT191127 from $z\gtrsim$2, but at such redshifts it is closer to the EP sample in duration. XRT100424 and XRT040610 never quite cross this region except for $z>6$.
From this Figure, we see that either the \citet{Quirola_Vasquez_2022, Quirola_Vasquez_2023}--\emph{Chandra} detected sample does not include all types of FXTs, or the FXTs from our sample originate from host galaxies at (much) higher redshifts than the ones we associated with them. In the latter case, the X-ray emission of XRT100424 would have to have travelled through the galaxy at $z=0.18$, because of its apparent position on top of this galaxy. As shown e.g.~ in \cite{jonker2010} this is unlikely as the X-rays would have been partially absorbed.

\subsubsection{Einstein Probe detected FXTs}
In Fig.~\ref{fig:lum_duration}, there seems to be only one EP source with a duration higher than 1ks and peak X-ray luminosity similar to those of the FXT sample of this paper ($10^{46}$~\lum), EP250108a. \cite{Eyles-Ferris2025} mention this low luminosity and long duration and suggest an intrinsically weaker or possibly failed jet as an explanation. Additionally we plot GRB~060218 \citep[SN~2006aj][]{Campana2006}, which is classified as an X-ray flash, and has a much longer duration and lower luminosity than most GRBs. 
EP-discovered FXTs mostly follow a luminosity relation with redshift \citep{Guo2025}. EP-discovered FXTs at $z<1$ have peak luminosities in the order of $10^{45} - 10^{48}$\lum, with more than half of this sample in \citep{Guo2025} having $10^{48}$\lum. EP-discovered FXTs at higher redshift ($z>1$) have peak luminosities of $10^{48}$~-~$10^{50}$~\lum~(this is a selection effect, as EP-WXT cannot detect any fainter FXT if at such a high redshift). 
From the FXTs in our sample (not EP-detected), only the peak X-ray luminosity of XRT191127 and Swift~J050400 if from D1 at $z=0.95$, would just fall within the lower end of these ranges. We calculate at what redshifts the other FXTs would have had to originate from in order to match the luminosities from the EP-discovered FXTs. For XRT100424, XRT140507 and XRT040610, it seems very unlikely that they originate from a galaxy at such a high redshift that its peak X-ray luminosity would match the ones detected by EP (i.e., for XRT100424, XRT140507, and XRT040610 a redshift of $z=32$, $z=18$, $z=36$ would be needed to reach $L_{X,peak}=10^{48}$\lum, respectively). Swift~J050400 would have to be at $z=12$ to reach $L_{X,peak}=10^{48}$\lum~and XRT151121 would have to be at $z=8$ to reach such a luminosity.
The rates of low-luminosity events in nature are higher than those with high luminosities. As \emph{Chandra} and XMM-\emph{Newton} have only a small field of view but high sensitivity, it is therefore more likely that they discover examples of the more numerous but lower luminosity events, whereas EP has a much larger field of view and can therefore also detect the rarer, brighter events. However, besides the luminosity also the rest-frame durations are different.
Alternatively, if the \emph{Chandra} and XMM-\emph{Newton}-discovered FXTs are not drawn from the low-luminosity tail of a $L_X$ distribution, the FXTs detected by \emph{Chandra} and XMM-\emph{Newton} could have (a) different origin(s) compared to the ones detected by EP. 
EP-discovered FXTs often have a duration in the order of a hundred seconds \citep[e.g.~][]{Wu2025}. From the FXTs not detected by EP in our sample, only XRT191127 has a similar duration ($T_{90}\approx300$~sec). The FXTs that have potential lower luminosities, also have much higher durations of 5.6~ks, 4.8~ks, 3.2~ks, 5.5~ks and 1.9~ks in observer frame for XRT100424, XRT140507, XRT040610, XRT151121 and Swift~J050400, respectively and in rest frame 4.7~ks, 4.4~ks, 1.7~ks, 1.6~ks and 970~s for the redshifts of the most likely candidate hosts we identified and $z=0.95$ for Swift~J05040 as it is the highest redshift for that FXT. 

Finally, we compare the isotropic energies ($E_{iso}$) of the FXTs. For XRT140507, XRT151121 and XRT191127 we use the $E_{iso}$ as calculated in \cite{Quirola_Vasquez_2023} and convert it to our newly-found redshifts. For the other FXTs, we take the average flux and the duration to find $E_{iso}$. We find that $E_{iso}$ for the EP-discovered FXTs is generally higher, with a mean of $E_{iso}\approx 10^{51}$~erg, a maximum of $E_{iso}\approx 10^{53}$~erg and a minimum of $E_{iso}\approx 10^{48}$~erg, for the FXTs considered. Whereas for the FXTs in our sample, not detected by EP, we find $E_{iso}\approx 10^{50}$~erg, $E_{iso}\approx 10^{49}$~erg and $E_{iso}\approx 10^{44}$~erg as maximum, mean and minimum values. 

\begin{table*}[ht]
    \caption{Summary of the progenitor scenarios for each FXT, assuming it belongs to the listed candidate host. }
    \label{tab:origins_summary}
    \centering
    \begin{tabular*}{\linewidth}{@{\extracolsep{\fill}}c|c|c|c|c|c}
       \multirow{2}{*}{FXT} & LGRBs  & \multirow{2}{*}{BNS merger} & \multirow{2}{*}{SN SBO} & \multicolumn{2}{c}{WD-IMBH TDE} \\ 
         &  (cocoon)&  &  & non-rel. & rel. \\ \hline
        XRT040610 A1& $L_{X,peak}< $ & \checkmark & \checkmark & $L_{X,peak}> $ &\checkmark\\
        XRT040610 A2& $L_{X,peak}< $& \checkmark & \checkmark & $L_{X,peak}> $ &\checkmark\\

        XRT040610 A3& $L_{X,peak}<$, Offset$> $ & $L_{X,peak}< $ & \checkmark & \checkmark & $L_{X,peak}< $\\ 
            
        XRT100424 & $L_{X,peak}< $, SFR/M$_*<$  & $L_{X,peak}< $ & \checkmark &  \checkmark & $L_{X,peak}< $\\ 
        
        XRT140507 B1& $L_{X,peak}< $, Offset$>$, SFR/M$_*<$ & $L_{X,peak}< $ & \checkmark & \checkmark & $L_{X,peak}< $\\ 

        XRT140507 B2& $L_{X,peak}< $, Offset$>$ & $L_{X,peak}< $ & \checkmark & \checkmark & $L_{X,peak}< $\\ 
        XRT151121 C1& \checkmark & \checkmark& $L_{X,peak}> $& $L_{X,peak}> $ &\checkmark\\ 
        XRT151121 C2& $L_{X,peak}< $, Offset$>$ & $L_{X,peak}< $, Offset$>$ & Offset$>$ & \checkmark & $L_{X,peak}< $ \\ 

        XRT191127 & $L_{X,peak}< $ & \checkmark & $L_{X,peak}> $ & $L_{X,peak}> $ & \checkmark\\ \hline
        Swift J050400 D1& $L_{X,peak}< $, Offset$>$, SFR/M$_*<$  & \checkmark & Offset$>$ & $L_{X,peak}> $ & \checkmark\\ 
        Swift J050400 D2& $L_{X,peak}< $ & \checkmark & \checkmark & $L_{X,peak}> $ & \checkmark\\  \hline
        EP240708a E1& SFR/M$_*<$ & \checkmark & $L_{X,peak}> $ & $L_{X,peak}> $&\checkmark \\
        EP240708a E2& \checkmark & \checkmark & $L_{X,peak}> $ & $L_{X,peak}> $ &\checkmark \\ \hline
    \end{tabular*}
    \tablefoot{Viable progenitor scenarios are indicated by a \checkmark. Viability is assessed based on three properties: $L_{X,peak}$, host galaxy offset and SFR/M$_*$; see the text for the constraints for each scenario. For non-viable progenitor scenarios, the violated constraint is listed. Values below or above the allowed ranges are denoted with $<$ and $>$, respectively.}
\end{table*}

\section{Conclusions}
\label{SectConclusion}
We present a study of the (search for the) host galaxies of eight FXTs. For five of these FXTs, Swift~J050400, XRT140507, XRT040610, XRT151121 and EP240708a, no candidate host galaxy has been reported in the literature before. We find at least one candidate host for each of these FXTs. 
Furthermore, we present optical spectra of (some of) the candidate host galaxies for XRT100424, XRT191223, Swift J050400 and EP240708a. 
We found the spectrum of the host for XRT191223 to be consistent with that of an M star. We conclude that this FXT is due to a flare from a Galactic M star and not an extra-galactic FXT.
We use photometric and spectroscopic data to perform SED fitting using the python fitting package BAGPIPES. This provides redshifts, galactic stellar masses, star formation rates, metallicities, ages and dust extinction for all the candidate host galaxies.

We use these properties, combined with the implied peak X-ray luminosity based on the determined redshift and the offset from the FXT to the galaxies to rule out certain proposed progenitor scenarios. 
A BNS merger origin is a plausible scenario for Swift J050400, XRT191127, XRT040610 and EP240708a. An SN SBO origin is a viable explanation for XRT100424, XRT140507 and XRT151121. A non-relativistic WD-IMBH can explain XRT100424, XRT140507, XRT040610 and XRT151121 and when relativistic, a WD-IMBH origin can explain Swift~J050400, XRT191127 and EP240708a. A collapsar as for LGRBs could be the progenitor for EP240708a or XRT151121. It is clear that we would need various origins to explain FXTs, if these galaxies are indeed the true hosts, but fainter host galaxies at higher redshifts cannot be ruled out.

We notice a difference in peak X-ray luminosities and durations between the FXTs detected with EP, those reported in previous studies of the (host galaxies of) \emph{Chandra}-discovered FXTs and those we discuss in this work. This could be explained by the possibility that the actual host galaxies of the FXTs in the sample discussed in this work are at a higher redshift, or that the two samples have a different origin.
The typical peak X-ray luminosities for EP-discovered FXTs is a few orders of magnitude higher than most FXTs in our sample. The duration of EP-discovered FXTs is $\sim0.1$~ks, compared to several ks for most FXTs in our sample. Only XRT151121 shares overlap in peak X-ray luminosity and XRT191127 in duration with the low-luminosity and long duration part of the EP-discovered FXTs. This could mean that EP is looking at a different type of FXTs with possibly other origins than the ones detected by \emph{Chandra} and XMM-\emph{Newton}. Alternatively, \emph{Chandra} and XMM-\emph{Newton} could be looking at the more numerous, low-luminosity events and EP at the rare, brighter events of the same population due to the difference in sensitivity and field of view between the satellites.

\section*{Data availability}
All spectra used in this work are available on WISeREP \citep{wiserep}. The Magellan-LDSS3 spectrum for XRT100424 can be found at \url{https://www.wiserep.org/object/30868}. The GTC/OSIRIS(+) spectra for XRT191223 at \url{https://www.wiserep.org/object/30869} and for Swift~J050400 at \url{https://www.wiserep.org/object/30870}. All spectra for EP240708a are available at \url{https://www.wiserep.org/object/30871}.

\begin{acknowledgements}
We are deeply grateful to Tom Marsh for developing the {\sc molly} software, one of his many contributions to advancing the field of compact objects. 

A.P.C.H., P.G.J., J.Q.V., J.N.D.D., J.S.S.~ 
are supported by the European Union (ERC, Starstruck, 101095973). Views and opinions expressed are however those of the author(s) only and do not necessarily reflect those of the European Union or the European Research Council Executive Agency. Neither the European Union nor the granting authority can be held responsible for them.

D.M.S.~acknowledges support through the Ram\'on y Cajal grant RYC2023-044941, funded by
MCIU/AEI/10.13039/501100011033 and FSE+. D.M.S.~and M.A.P.T.~also acknowledge support by the Spanish Ministry of Science via the Plan de Generacion de conocimiento PID2021-124879NB-I00.

Based on observations obtained under the International Time Programme of the CCI (International Scientific Committee of the Observatorios de Canarias of the IAC) under program ID 23ITP (PI Jonker)
and 21A, 22A and 23B (PI Torres) with the GTC operated on the island of La Palma by the Roque de los Muchachos. HiPERCAM was funded by the European Research Council under the European Union’s Seventh Framework Programme (FP/2007-2013) under ERC-2013-ADG Grant Agreement no. 340040 (HiPERCAM), with additional funding for operations and enhancements provided by the UK Science and Technology Facilities Council (STFC).

Based on observations collected at the European Organisation for Astronomical Research in the Southern Hemisphere under ESO programme 113.26ET.004.

Based on observations with the Southern Astrophysical Research Telescope under program SOAR2024A-012, PI Bauer.

Based on observations obtained at the international Gemini Observatory (Program IDs GS-2024A-FT-114), a program of NOIRLab, which is managed by the Association of Universities for Research in Astronomy (AURA) under a cooperative agreement with the National Science Foundation on behalf of the Gemini Observatory partnership: the National Science Foundation (United States), National Research Council (Canada), Agencia Nacional de Investigaci\'{o}n y Desarrollo (Chile), Ministerio de Ciencia, Tecnolog\'{i}a e Innovaci\'{o}n (Argentina), Minist\'{e}rio da Ci\^{e}ncia, Tecnologia, Inova\c{c}\~{o}es e Comunica\c{c}\~{o}es (Brazil), and Korea Astronomy and Space Science Institute (Republic of Korea). Data was processed using the Gemini DRAGONS (Data Reduction for Astronomy from Gemini Observatory North and South) package.
\end{acknowledgements}

\bibliographystyle{aa}
\bibliography{bibliography}

\begin{appendix}
\section{Sample of FXTs}
\label{AppendixSample}
We present a short summary of the properties of the eight FXTs in Table~\ref{table:sample}.

\begin{table*}[ht!]
\caption{The sample of FXTs.}
\label{table:sample}      
\centering 
\begin{tabular*}{\linewidth}{@{\extracolsep{\fill}}c c c c c c}   
\hline\hline \\
 Name & RA (degrees) & Dec (degrees) & Pos. Unc. (arcsec) & F$_{X,peak}$ (erg s$^{-1} \mathrm{cm}^{-2}$) &  ref \\ \hline \\
        
        XRT040610 & 169.53625 & 7.70266& 3.0 & (0.45$\pm$0.08)$\times$10$^{-13}$ & (1) \\
        
        XRT100424 & 321.79670 & -12.03906 & 3.4 
        & (0.57$\pm$0.12)$\times$10$^{-13}$ & (1) \\ 

        XRT140507 & 233.73496 & 23.46849 & 0.8 & (1.9$\pm$0.9)$\times$10$^{-13}$ & (2) \\

        XRT151121 & 40.82972 & 32.32390 & 2.0 & (1.2$\pm$0.5)$\times$10$^{-12}$ & (2) \\

        XRT191127 & 207.34711 & 26.58421 & 1.2 & (1.5$\pm$0.6)$\times$10$^{-11}$ & (2)\\
        
        XRT191223 & 50.47516 & 41.24704 & 0.4 & (1.9$\pm$0.9)$\times$10$^{-12}$ & (2)  \\ 

        Swift J050400 & 76.0014 & 67.5680 & 4.9 & (4.2$^{+7.3}_{-2.5}) \times 10^{-13}$ &  (3) \\
        EP240708a & 345.9641 & -22.8430 & 9.3 & $\sim$1.1$\times$10$^{-9}$ & (4) \\
\hline            
\end{tabular*}
\tablefoot{The peak X-ray flux is reported in the 0.3-10 keV energy range, except for EP240708a where it is in the 0.5-4 keV energy range. The positional uncertainty is given at the 2$\sigma$ confidence level.}
\tablebib{(1)~\citet{AlpLarsson2020}; (2) \citet{Quirola_Vasquez_2023}; (3) \citet{Evans2023}; (4) \citet{2024GCNEP240708a}.}
\end{table*}

\section{Observations and data reduction}
\label{app:obs+datared}

The Gran Telescopio Canarias (GTC), located at the Roque de los Muchachos observatory on La Palma, Spain, observed the candidate host galaxies of XRT100424, XRT1911123 and Swift~J050400 with the Optical System for Imaging and low-Intermediate-Resolution Integrated Spectroscopy \citep[OSIRIS][named OSIRIS+ for data obtained after January 2023 due to a major upgrade of the instrument]{cepa1998osiris} and the fields of XRT140507, XRT040610 and XRT151121 with HiPERCAM \citep{2016Dhillon,2018Dhillon,2021Dhillon}.

We obtained long-slit spectra of the candidate hosts of XRT100424 and XRT191223 identified in the literature (Section \ref{SectSample}) and of two galaxies identified by us (Section~\ref{SectResults}) of Swift~J050400. All spectra were taken with a slit-width of 1" and grism R500R. This setting allows to cover the wavelength range 4800~-~10000~\AA\ and has a resolution power of R~=~587 at the central wavelength of $\lambda_c = 7165$~\AA. The spectra were reduced and flux calibrated with spectro-photometric standard stars observed at the end of the night. For this calibration we either used dedicated commands in PypeIt \citep{pypeit} or in {\sc molly} \citep{Molly}.

We observed the candidate host galaxy of XRT100424 using the Magellan II Clay telescope, located at Las Campanas Observatory, Chile, with the Low Dispersion Survey Spectrograph (LDSS-3). We obtained a long-slit spectrum of the host galaxy of XRT100424. We used the VPH-All grism, which has wavelength coverage of 4250~-~10000~\AA\, with a central wavelength of 7100~\AA. The spectrum was wavelength and flux calibrated using {\sc molly} \citep{Molly}.

Images in \textit{i-}band of the field of EP240708a were taken on two epochs with the Southern Astrophysical Research (SOAR) Telescope, located at Cerro Pach\'on in Chili with the Goodman High Throughput Spectrograph (GHTS, \cite{Clemens2004}). A difference image of the two was made by using \texttt{PyZOGY} \citep{Guevel2017} to search for a possible counterpart of EP240708a, but none was found.

The Very Large Telescope (VLT) observed the fields of XRT140507 and XRT191127 using the FOcal Reducer and low dispersion Spectrograph 2 \citep[FORS2,][]{fors} under Program ID 111.24P0.009, and the candidate host galaxy of EP240708a using X-Shooter \citep{xshooter} with Program ID 113.26ET.004. 

\label{xshooter}
A spectrum was taken with X-SHOOTER of the candidate host galaxy of EP240708a marked as E2 in Fig.~\ref{fig:finders} with 6x300~s exposures in the NIR arm and single exposure of 1200~s in UVB and VIS arm. The spectrum was reduced using the ESO X-shooter workflow in ESOReflex \citep{ESOReflex}.

\label{gemini}
The candidate host galaxy EP240708a E1 was observed by the Gemini South Telescope using the GMOS-S instrument. We obtained a spectrum of the host galaxy E1 on 29~September~2024, with a total on-target exposure time of $4\times320$~sec. Our spectroscopic observations were all carried out using the R150 grating, and a 1\arcsec slit width. The spectra cover the wavelength range $\approx$450–1000~\AA. We reduced GMOS data using the {\sc DRAGONS} pipeline \citep{Labrie2023a,Labrie2023b} and followed standard recipes for the spectroscopic observation.

\begin{table*}[ht!]
\caption{A journal of the new photometric and spectroscopic observations of the host candidates of the FXTs obtained for this work. 
}
\label{table:obsjournal}      
\centering 
\begin{tabular*}{\linewidth}{@{\extracolsep{\fill}}c c c c c c c} 
\hline\hline
 Target & Telescope/ & Date & Filter / grism & T$_{\rm exp}$ & Seeing & Air mass \\ 
   & Instrument &  &  &  (\# $\times$ s) &(") & \\
 \hline  
  XRT040610 & GTC/HiPERCAM & 2024 Feb 11 & \textit{u,g,r,i,z} & 30 $\times$ 100 & 0.6 & 1.07 \\
        \hline
        
      XRT100424 & GTC/OSIRIS & 2021 Aug 5 & R500R & 4~x~600 & 1.2 & 1.3 \\
    XRT100424 & Magellan/LDSS-3 & 2021 Oct 12 & VPH-all & 1200 & 1.0 & 1.3 \\ \hline 

        XRT140507 & VLT/FORS2 & 2023 Jul 14& U, B, V, R, I & 6 $\times$ 120 & 0.61 - 0.81
        & 1.6 \\
        XRT140507 & VLT/FORS2 & 2023 Jul 15 & U, B, V, R, I & 6 $\times$ 120 & 0.49 - 0.66 & 1.6 - 1.7\\
        XRT140507 & VLT/FORS2 & 2023 Jul 19 & U, B, V, R, I & 12 $\times$ 120 & 0.38 - 0.65
        & 1.6 - 2.1 \\
        XRT140507 & VLT/FORS2 & 2023 Jul 26 & U, B, V, R, I & 12 $\times$ 120 & 0.38 - 0.54 
        & 1.6 - 2.2 \\
        XRT140507 & VLT/FORS2 & 2023 Aug 3 & U, B, V, R, I & 6 $\times$ 120 & 1.04 - 1.46  &  1.5 - 1.6\\
        XRT140507 & GTC/HiPERCAM & 2024 Feb 11 & \textit{u,g,r,i,z} & 23 $\times$ 100 & 0.6 & 1.3\\
        \hline

        XRT151121  & GTC/HiPERCAM & 2024 Feb 11 & \textit{u,g,r,i,z} & 30 $\times$ 100 & 0.8& 1.10 \\
        \hline

           XRT191127 & VLT/FORS2 & 2023 Jun 27 & U,B,V,R,I & (6+6+8+6+6)  $\times$ 120 & 0.87 - 1.69 & 1.7 - 2.0\\
        XRT191127 & VLT/FORS2 & 2023 Jun 28 & U,B,V,R,I & 6  $\times$ 120 & 0.77 - 1.69 & 1.7 - 2.0\\
        XRT191127 & VLT/FORS2 & 2023 Jul 15 & U,B,V,R,I & 18 $\times$ 120 & 0.47 - 0.85 & 1.6 - 2.3\\
        XRT191127 & VLT/FORS2 & 2023 Jul 19 & U,B,V,R,I & 6 $\times$ 120 & 0.37 - 0.50 & 1.6 - 1.7\\
        XRT191127 & VLT/FORS2 & 2023 Jul 21 & U,B,V,R,I & (8+6+6+6+6)  $\times$ 120  & 0.55 - 0.75 & 1.7 - 1.8\\
        XRT191127 & VLT/FORS2 & 2023 Jul 24 & U,B,V,R,I & (6+6+6+8+6) $\times$ 120 & 0.66 - 0.96 & 1.6 - 1.8 \\
        XRT191127 & VLT/FORS2 & 2023 Jul 26 & U,B,V,R,I & (12+6+6+8+6)  $\times$ 120 & 0.52 - 0.67 & 1.9 - 2.6\\
        \hline
    
            XRT191223 & GTC/OSIRIS & 2022 Aug 24, 26 & R500R  & 2607 & 1.1, 1.2 & 1.1 \\ \hline 
        
               Swift~J050400 & GTC/OSIRIS+ & 2023 Dec 14 &  $g^\prime$, $i^\prime$  & 3 $\times$ 200 & 2.0, 1.3 & 1.3 \\
     Swift~J050400 & GTC/OSIRIS+ & 2023 Dec 15 & $i^\prime$  & 6 $\times$200 & 0.7 & 1.3 \\
     Swift~J050400 & GTC/OSIRIS+ & 2024 Jan 2 & $i^\prime$ & 6 $\times$ 200 & 0.8 & 1.3 \\
     Swift~J050400 & GTC/OSIRIS+ & 2024 Feb 2 & R500R  & 1200 &1.2, 0.7 & 1.3 \\ 
        \hline        
        EP240708a & SOAR/GHTS & 2024 Jul 9 & i & 180 & 1.2 & 1.0 \\
        EP240708a & SOAR/GHTS & 2024 Aug 10 & i &  240 & 2.0 & 1.0 \\
        EP240708a & VLT/X-SHOOTER & 2024 Sept 30 & - & \begin{tabular}{@{}c@{}}1200, 1200,\\ 6x300\end{tabular}& 0.50-0.76 & 1.1\\ 
        EP240708a & Gemini South/GMOS & 2024 Sept 29 & R150 & 4 $\times$ 320 & 0.9 & 1.1 \\ 
\hline                  
\end{tabular*}
\tablefoot{The name of the target is in the first column and the telescope and instrument used in the second.
The third column lists the date at the start of the night. 
The fourth column shows the filters for the photometric observations or the grism for the spectra, all spectra are taken with slit width 1\arcsec, expect for the Magellan/LDSS-3 and X-Shooter spectra where we used a 1.5\arcsec and 0.9\arcsec~slit width, respectively.
The exposure time in column five (T$_{\rm exp}$) is given as the number of exposures per filter (\#) times the time per exposure in seconds.
The seeing and airmass at the time of the observations are given in columns six and seven, respectively. In the case of VLT/FORS2 data, the ranges for the seeing and airmass over the duration of the observations and over all the filters are given.}
\end{table*}

\section{Additional finder charts}
\label{Section:AppendixFinders}
We provide a finder chart of XRT100424, for which the host galaxy had already been reported in the literature \citep{AlpLarsson2020}. We also provide the stack of all the $ugriz$ HiPERCAM filters of the field of XRT040610.

\begin{figure*}[ht]
        \centering
        \begin{subfigure}{0.49\linewidth}
    \centering
    \includegraphics[width=0.8\textwidth]{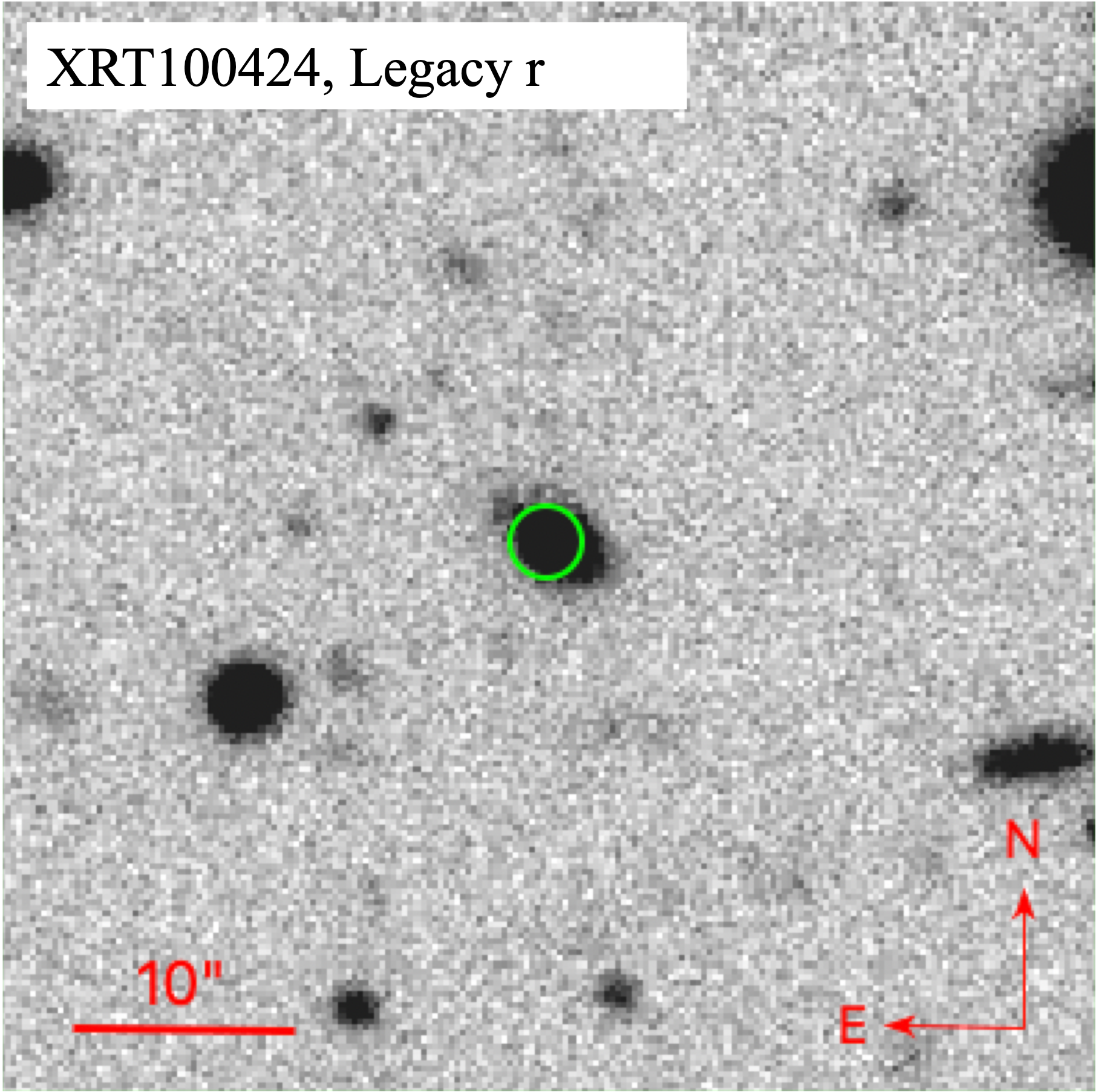}
    \subcaption{}
    \label{Fig:Finderxrt100424}
    \end{subfigure}
    \begin{subfigure}{0.49\linewidth}
    \centering
    {\includegraphics[width=0.8\textwidth]{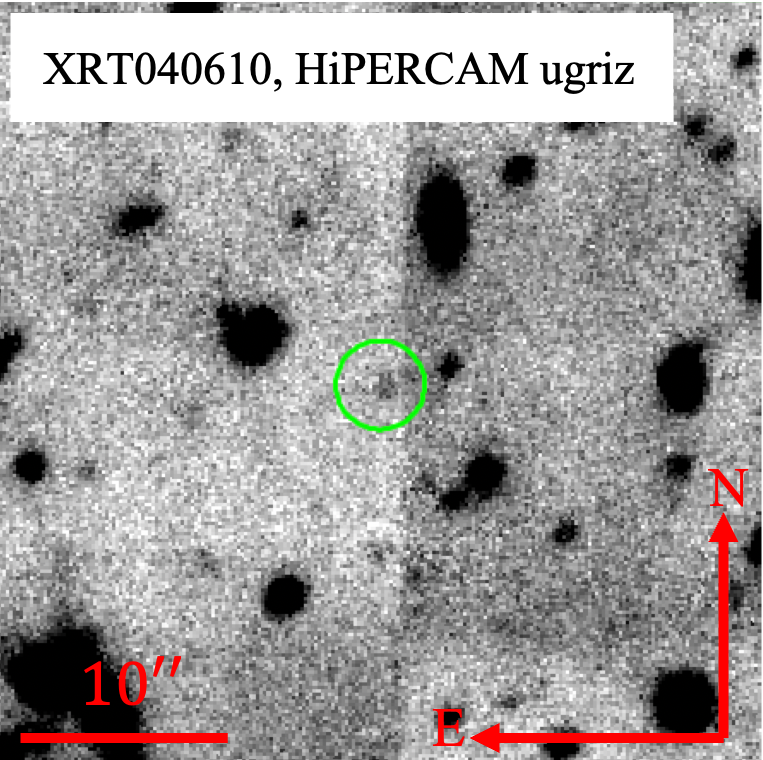}}
    \caption{}
    \label{fig:xrt040610_ugriz}
\end{subfigure}
    \caption{Left:) Finder chart of XRT100424 for which a candidate host galaxy was reported in the literature \citep{AlpLarsson2020}. The image is taken from the Legacy Survey \citep{Legacy}. The green circle denotes the 3$\sigma$ uncertainty of the X-ray position. (Right:) Stack of all filters $ugriz$ of the HiPERCAM observation of XRT040610. 
    }

\end{figure*}

\section{Limits on underlying galaxies}
\begin{figure*}[ht]
        \centering
        \begin{subfigure}{0.49\linewidth}
    \centering
    \includegraphics[width=\textwidth]{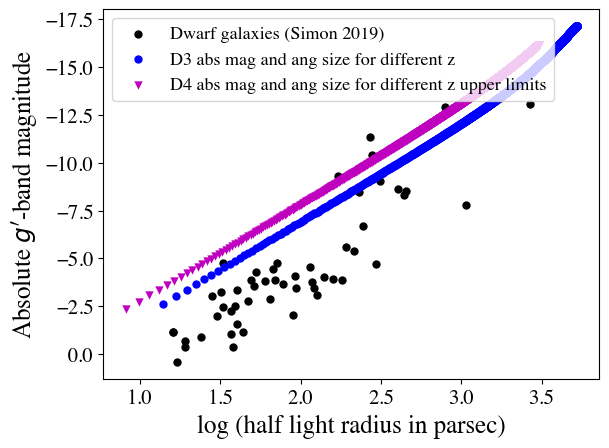}
         \subcaption{}
             \label{fig:Swift_D3D4_dwarfs}
    \end{subfigure}
    \begin{subfigure}{0.49\linewidth}
    \centering
    {\includegraphics[width=\textwidth]{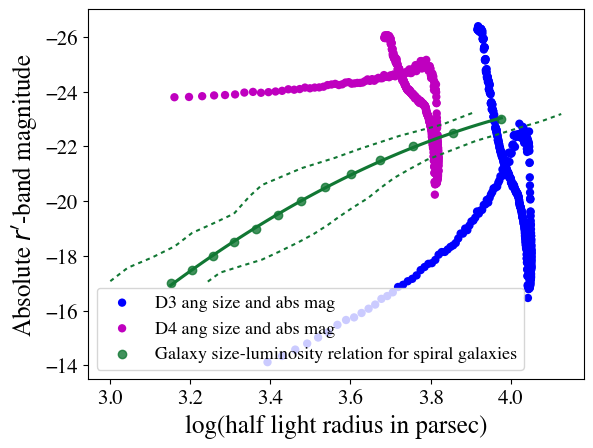}}
         \subcaption{}
             \label{fig:Swift_D3D4_spirals}
    \end{subfigure}
    
    \caption{The half-light radius of Swift~J050400 D3 (blue) and Swift~J050400 D4 (magenta) vs the $g'$-band magnitude compared to that of dwarf galaxies (black) from \cite{Simon2019} in the left panel, and the size-luminosity relation of spiral galaxies from \cite{zhang2019} (green) with the 1$\sigma$ uncertainty, in the right panel. The $g'$-band magnitudes of D4 are given as 3$\sigma$ upper limits, and could therefore be consistent with a dwarf galaxy at a large redshift range. 
    Swift~J050400 D3 is over all above the size-luminosity relation for dwarfs, except between $log(R_h)\sim2.2$ and $log(R_h)\sim2.5$. The relation for Swift~J050400 D3 is consistent with the spiral galaxy size-luminosity relation for redshifts in the ranges of 0.7-1.3 or 3.2-3.9. For Swift~J050400 D4 this range is 1.4-2.3. 
    }
    \label{fig:Swift_D3D4}

\end{figure*}

For Swift~J050400~D3 and Swift~J050400~D4, we use the (3$\sigma$ upper limits on the) absolute $g$-band magnitudes and half-light radii in parsec for different redshifts to see if they fall on either the size-luminosity relation for dwarf galaxies \citep[][]{Simon2019, Eappachen_2022} or spiral galaxies \citep{zhang2019}. They are shown in the left and right panels of Fig.~\ref{fig:Swift_D3D4}, respectively.

\section{SED fitting}
\label{Section:AppendixSEDfit}
The best-fit BAGPIPES fits with their posterior distributions of the fitted parameters are shown in Fig.~\ref{fig:allfits}.

\begin{figure*}[ht]
    \centering
    \begin{subfigure}{0.49\linewidth}
    \centering
    \includegraphics[width=\linewidth]{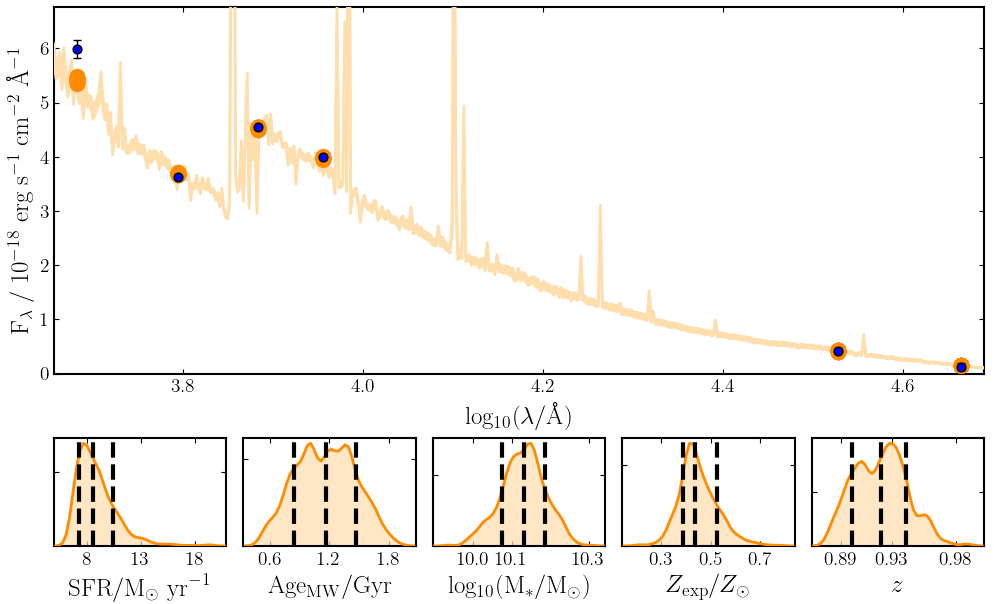}
         \subcaption{Best fit model of the HiPERCAM and unWISE photometry of XRT040610 A1.}
         \label{bagpipes:xrt040610_D1}
    \end{subfigure}
   \begin{subfigure}{0.49\linewidth}
   \centering
   \includegraphics[width=\linewidth]{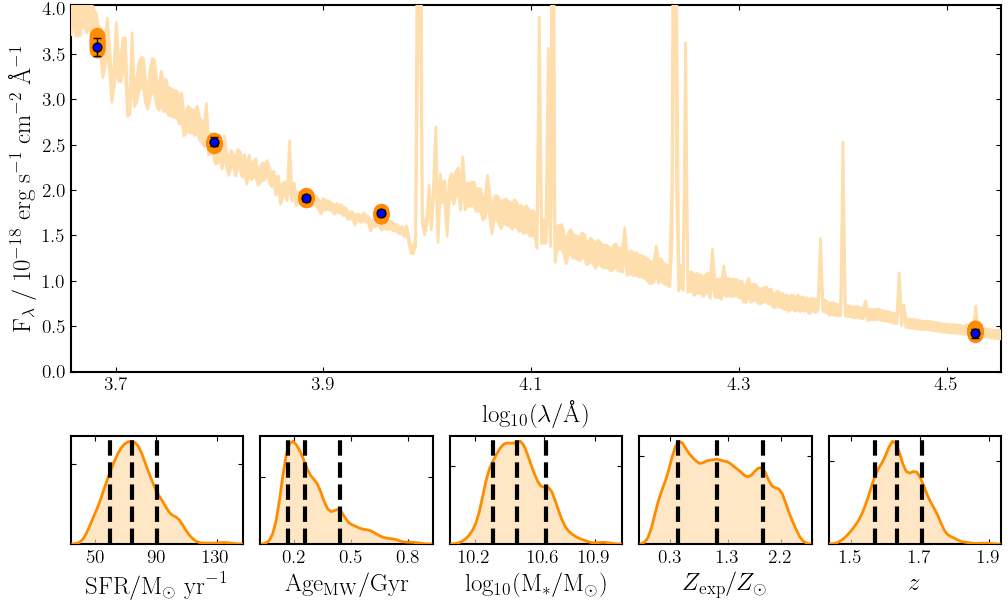}
         \subcaption{Best fit model of the HiPERCAM and unWISE photometry of XRT040610 A2.}
         \label{bagpipes:xrt040610_D2}
   \end{subfigure}
    \caption{
    Top panels: The best fitting models obtained using the BAGPIPES package are shown in orange. The data is shown in blue, together with the 1$\sigma$ uncertainties. For the spectra, the gray shaded areas are excluded from the fit as they cover wavelength regions of known sky emission lines and telluric absorption and the most important absorption (red) and emission (blue) lines for the fitted redshift are shown indicated by vertical dashed lines. 
    The wavelengths on the horizontal axes are in the observer frame. 
    Bottom panels: Posterior probability distributions for the five fitted parameters (SFR, age, galaxy stellar mass, metallicity and redshift). The 16th, 50th and 84th percentile posterior values are indicated by the vertical dashed black lines in each subplot.
    Panel a: HiPERCAM and unWISE photometry of XRT040610 A1. 
    Panel b: HiPERCAM and unWISE photometry of XRT040610 A2.
    Panel c: HiPERCAM and unWISE photometry of XRT040610 A3
    Panel d: Magellan/LDSS3 spectrum of the host galaxy of XRT100424.
    Panel e: SDSS-DR18 spectrum of XRT140507 B1.
    Panel f: SDSS-DR18 spectrum of XRT140507 B2.
    Panel g: HiPERCAM photometry of XRT151121 C1.
    Panel h: HiPERCAM and unWISE photometry of XRT151121 C2.
    Panel i: FORS2 and unWISE photometry of the host galaxy of XRT191127.
    Panel j: GTC/OSIRIS+ spectrum of Swift~J050400~D1, assuming the emission line is H$\alpha$ at $z=0.11$. 
    Panel k: GTC/OSIRIS+ spectrum of Swift~J050400~D1, assuming the emission line is [OIII] at $z=0.45$.
    Panel l: GTC/OSIRIS+ spectrum of Swift~J050400~D1, assuming the emission line is H$\beta$ at $z=0.49$. 
    Panel m: GTC/OSIRIS+ spectrum of Swift~J050400~D1, assuming the emission line is [OII] at $z=0.95$.
    Panel n: GTC/OSIRIS+ spectrum of Swift~J050400~D2.
    Panel o: GS spectrum of EP240708a~E1.
    Panel p: X-Shooter spectrum of EP240708a~E2.}
    \label{fig:allfits}
\end{figure*}

\begin{figure*}[ht]\ContinuedFloat
\centering
\begin{subfigure}{0.49\linewidth}
   \centering
        \includegraphics[width=\linewidth]{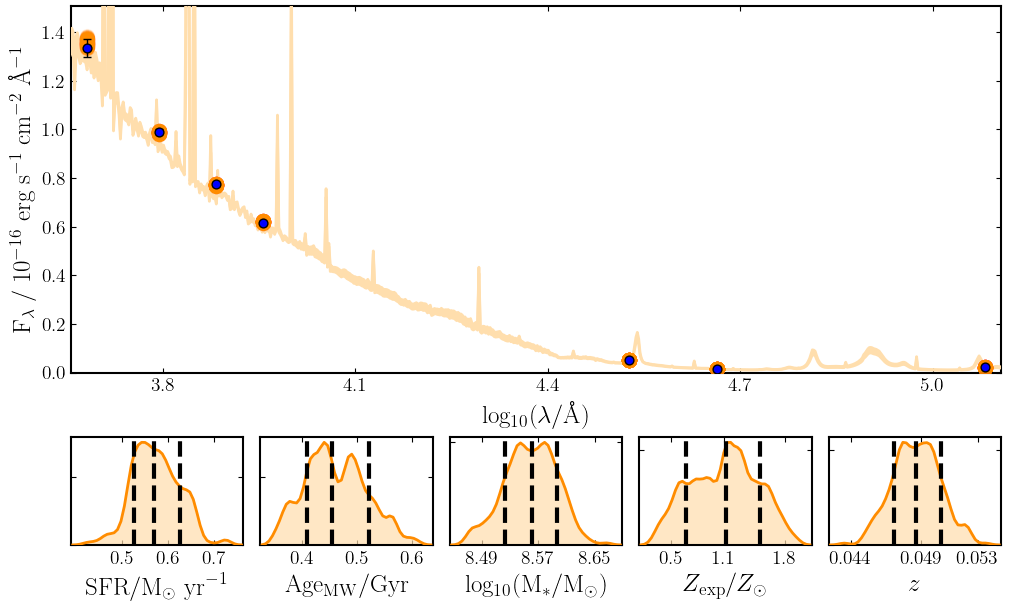}
         \subcaption{}
         \label{bagpipes:xrt040610_D3}
   \end{subfigure}
    \begin{subfigure}{0.49\linewidth}
    \centering
         \includegraphics[width=\linewidth]{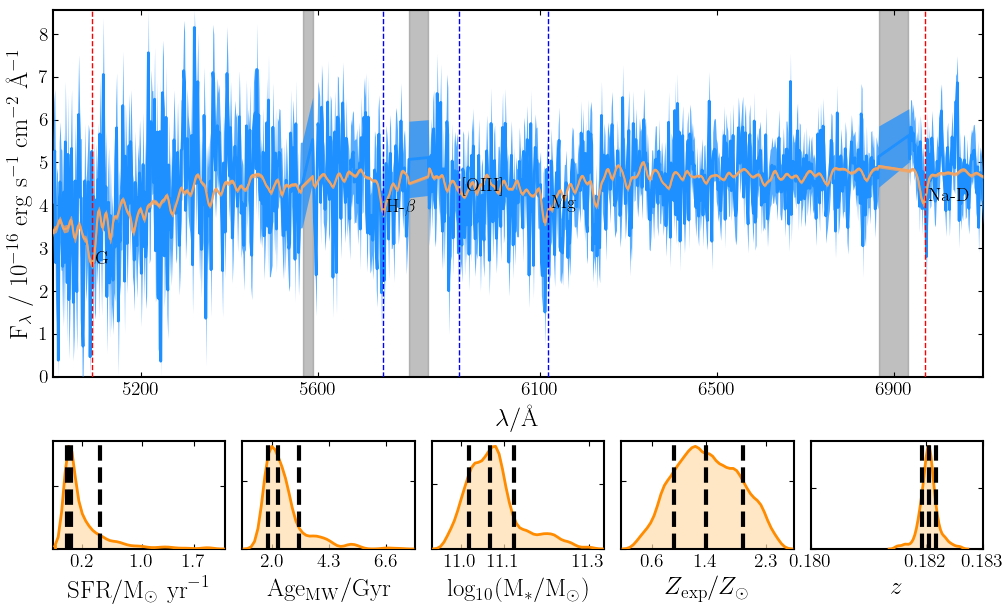}
         \subcaption{}
         \label{bagpipes:xrt100424}
    \end{subfigure}

    \begin{subfigure}{0.49\linewidth}
    \centering
         \includegraphics[width=\linewidth]{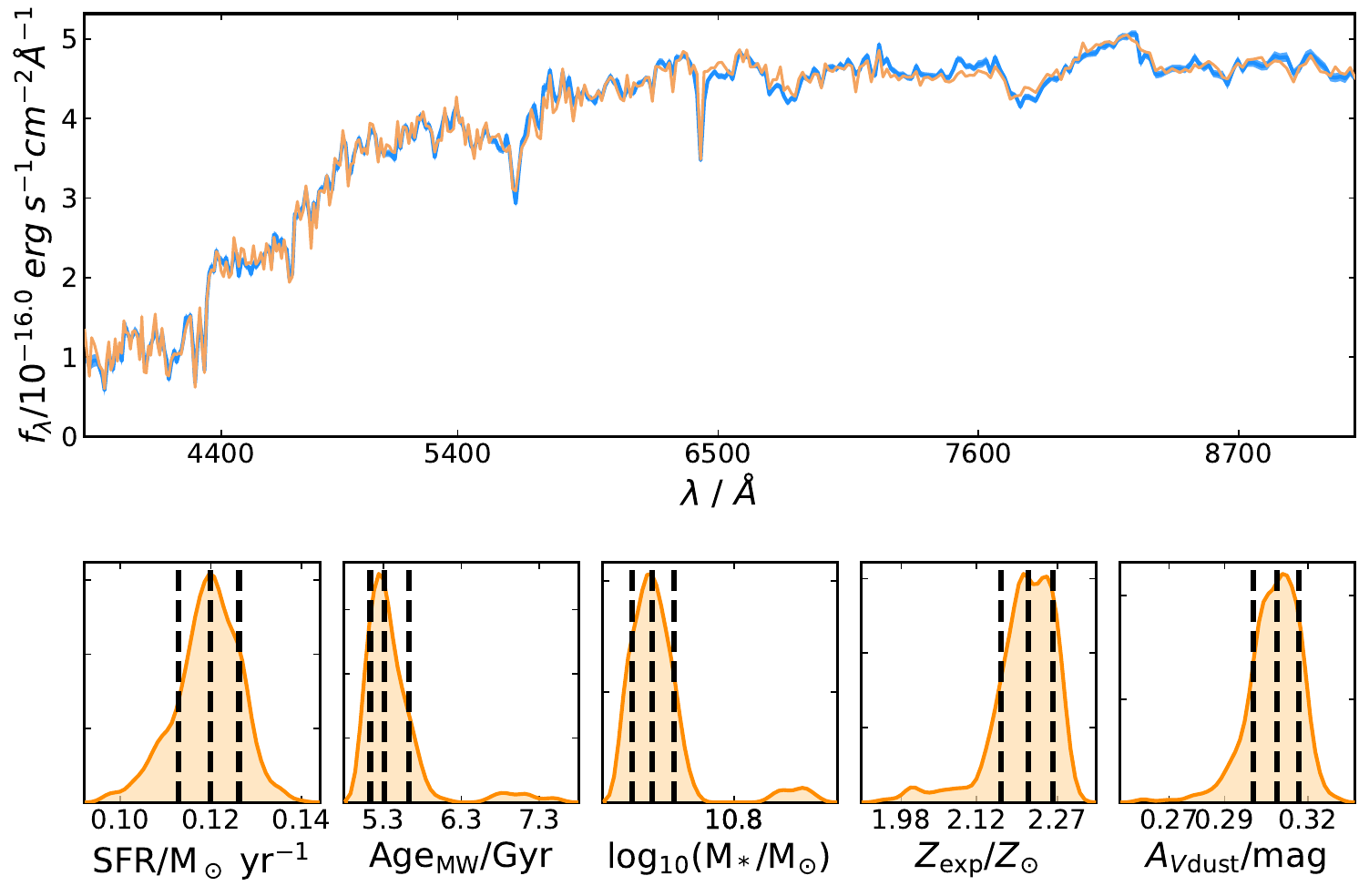}
         \subcaption{}
         \label{bagpipes:xrt140507_k1}
    \end{subfigure}
   \begin{subfigure}{0.49\linewidth}
   \centering
       \includegraphics[width=\linewidth]{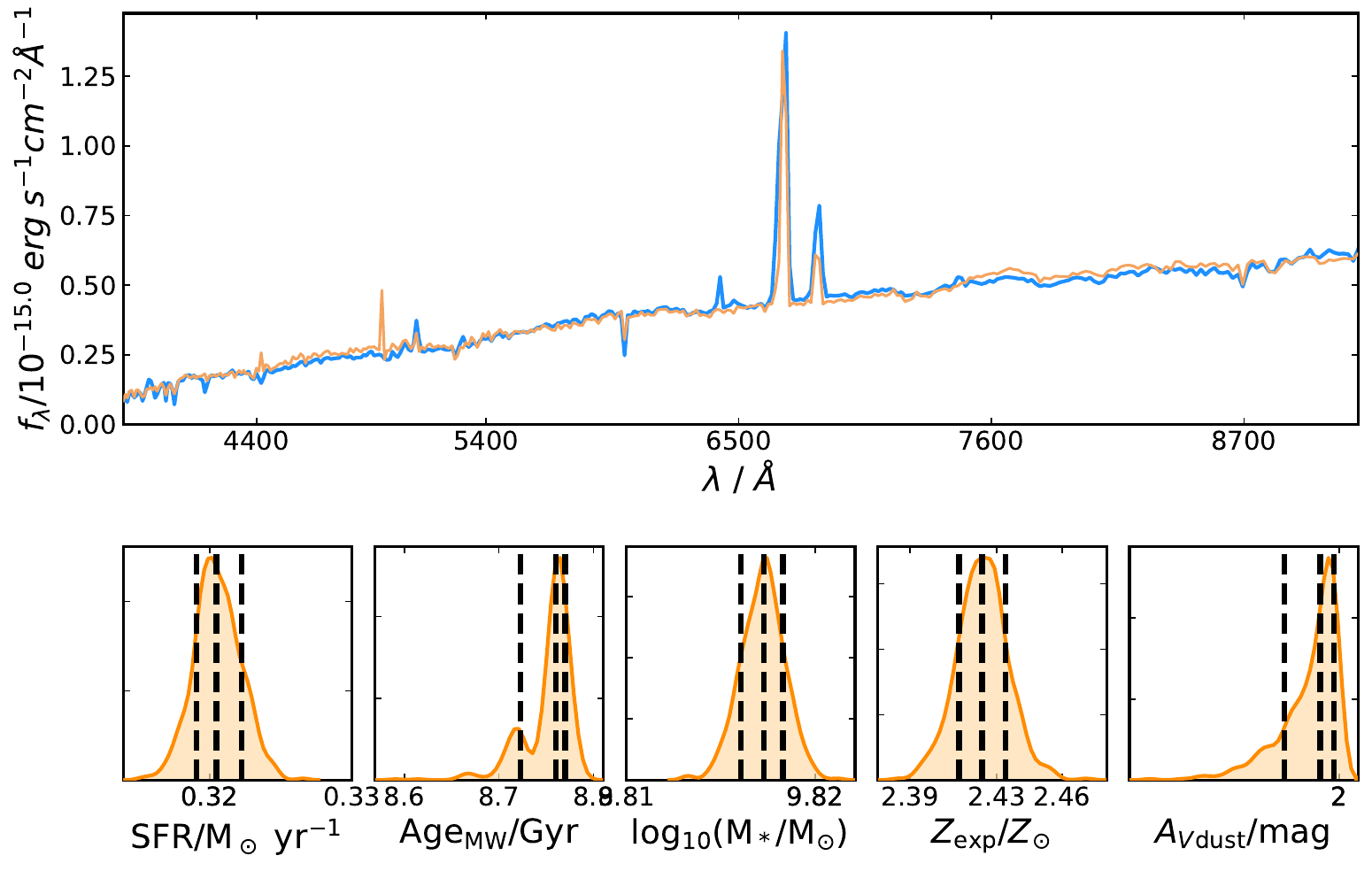}
         \subcaption{}
         \label{bagpipes:xrt140507_k2}
   \end{subfigure}

    \begin{subfigure}{0.49\linewidth}
    \centering
         \includegraphics[width=\linewidth]{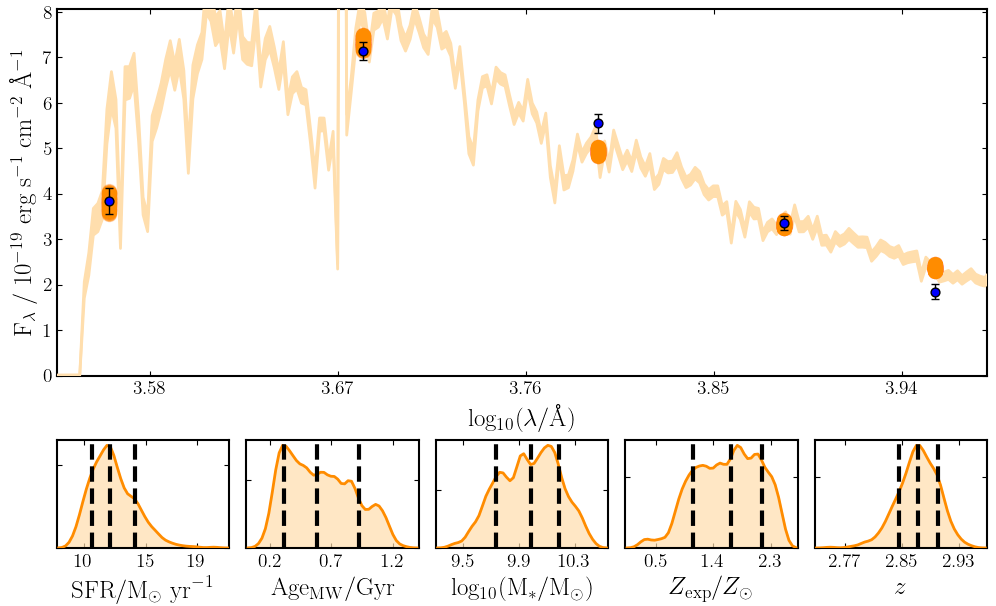}
         \subcaption{}
         \label{bagpipes:xrt151121_c1}
    \end{subfigure}
   \begin{subfigure}{0.49\linewidth}
   \centering
       \includegraphics[width=\linewidth]{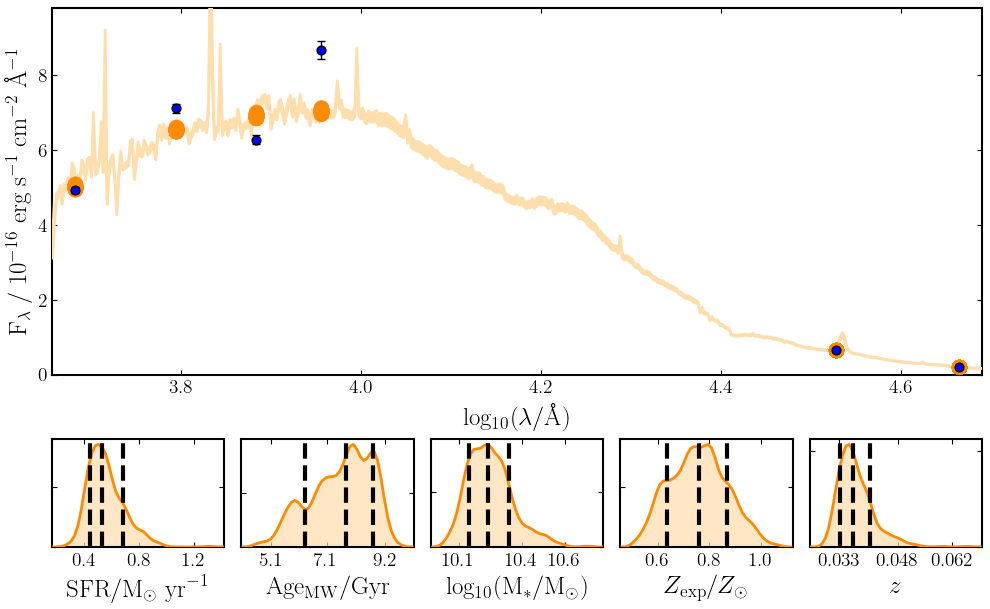}
         \subcaption{}
         \label{bagpipes:xrt151121_c2}
   \end{subfigure}
   \begin{subfigure}{0.49\linewidth}
    \centering
         \includegraphics[width=\linewidth]{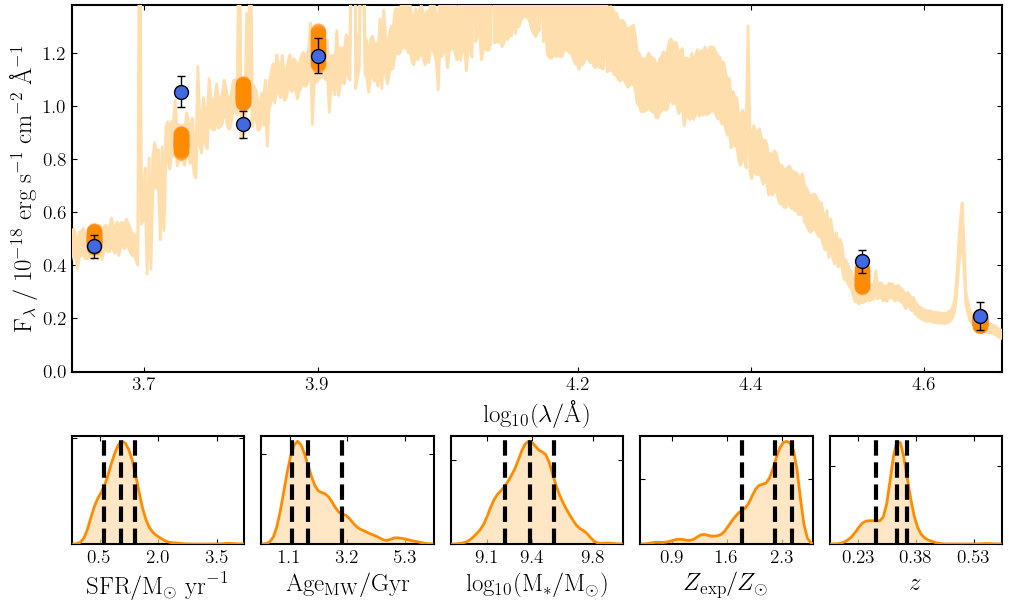}
         \subcaption{}
         \label{bagpipes:xrt191127}
    \end{subfigure}
    \begin{subfigure}{0.49\linewidth}
    \centering
         \includegraphics[width=\linewidth]{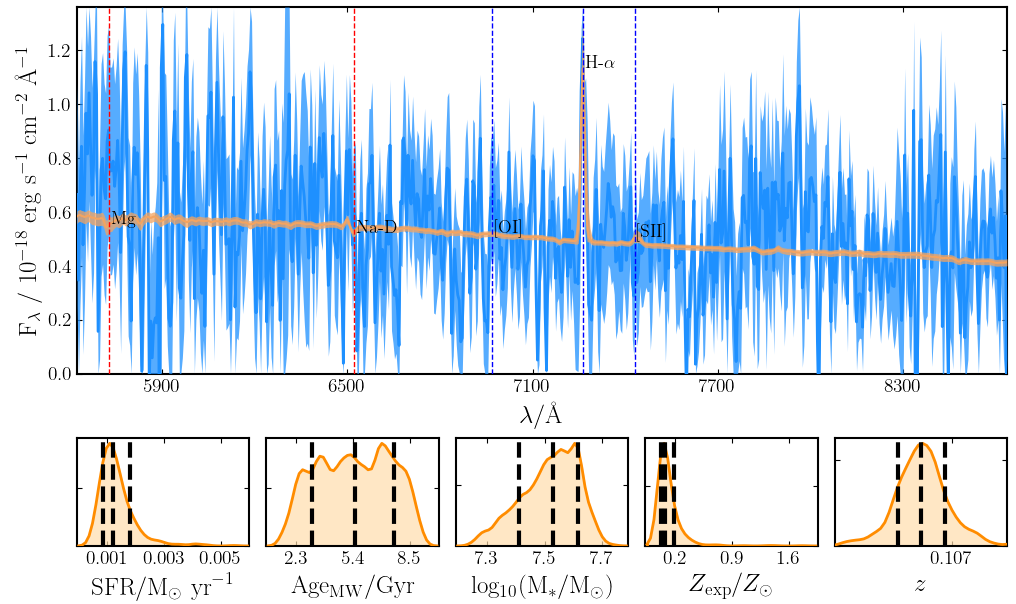}
         \subcaption{}
         \label{bagpipes:SwiftJ050400G2_z011}
    \end{subfigure}
    \caption{Continued.}
\end{figure*}

\begin{figure*}[ht]\ContinuedFloat
    \centering

   \begin{subfigure}{0.49\linewidth}
   \centering
       \includegraphics[width=\linewidth]{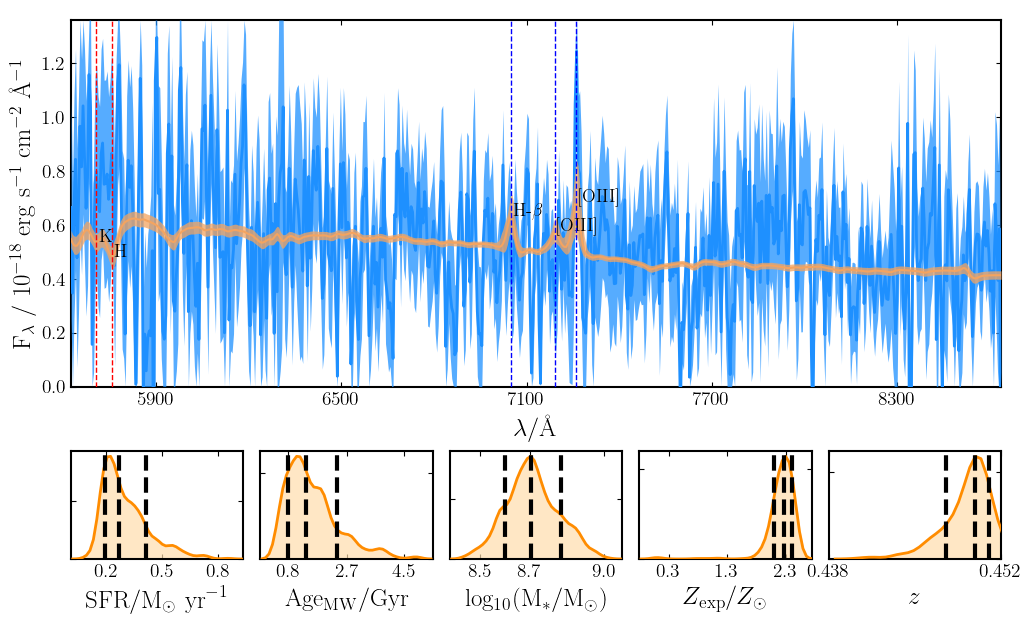}
         \subcaption{}
         \label{bagpipes:SwiftJ050400G2_z044}
   \end{subfigure}
   \begin{subfigure}{0.49\linewidth}
    \centering
         \includegraphics[width=\linewidth]{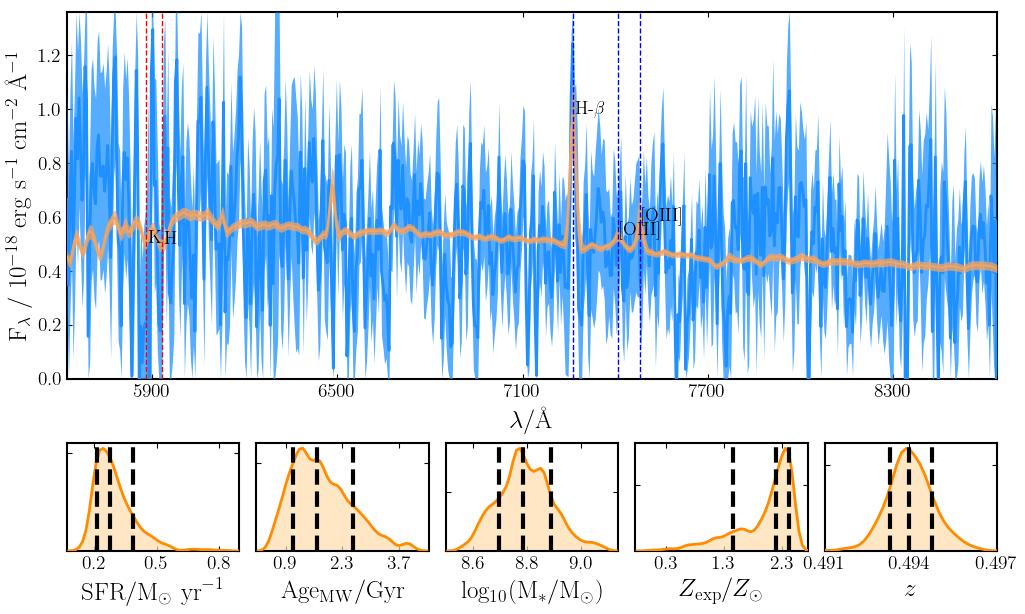}
         \subcaption{}
         \label{bagpipes:SwiftJ050400G2_z049}
    \end{subfigure}
   \begin{subfigure}{0.49\linewidth}
   \centering
       \includegraphics[width=\linewidth]{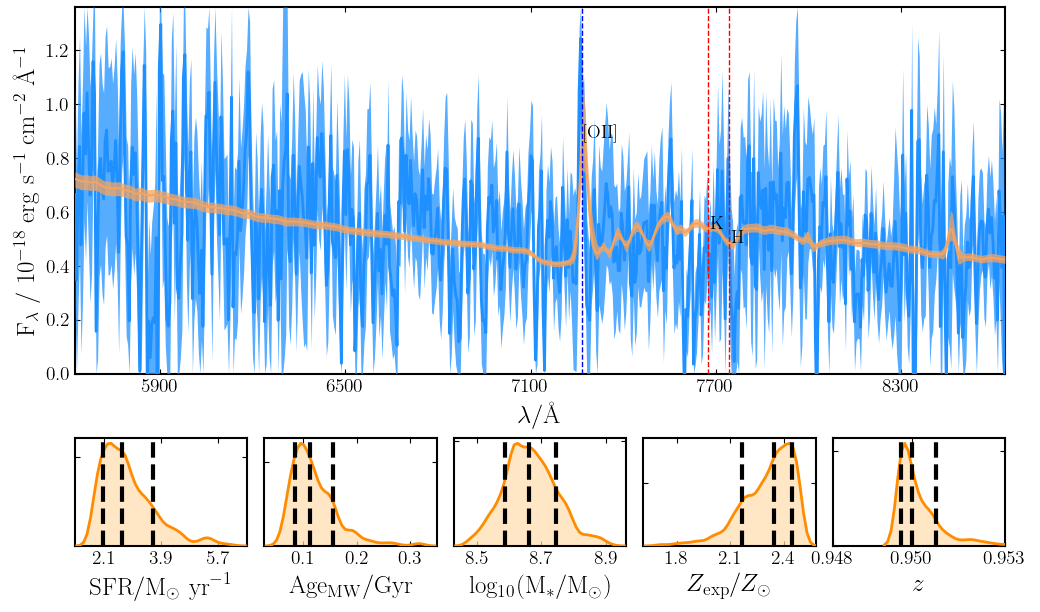}
         \subcaption{}
         \label{bagpipes:SwiftJ050400G2_z095}
   \end{subfigure}
   \begin{subfigure}{0.49\linewidth}
   \centering
       \includegraphics[width=\linewidth]{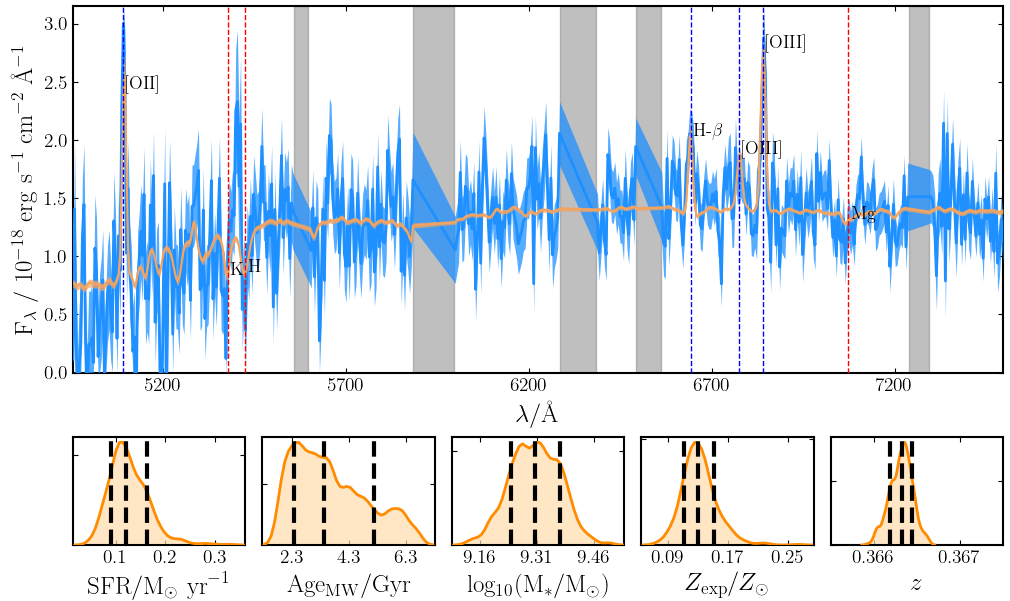}
         \subcaption{}
         \label{bagpipes:SwiftJ050400G1}
   \end{subfigure}

    \begin{subfigure}{0.49\linewidth}
    \centering
         \includegraphics[width=\linewidth]{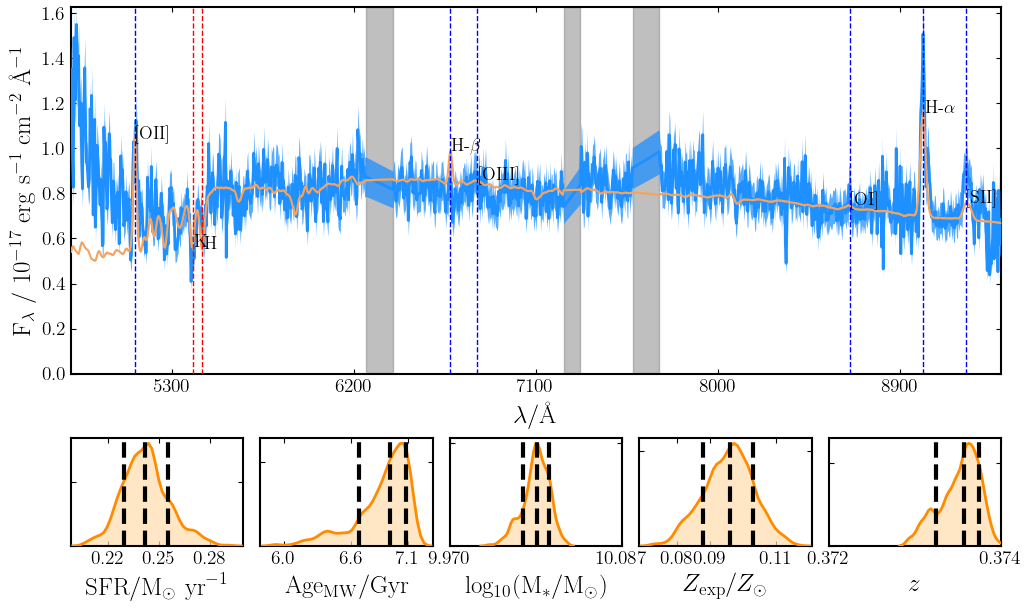}
         \subcaption{}
         \label{bagpipes:ep240708a_H1}
    \end{subfigure}
   \begin{subfigure}{0.49\linewidth}
   \centering
       \includegraphics[width=\linewidth]{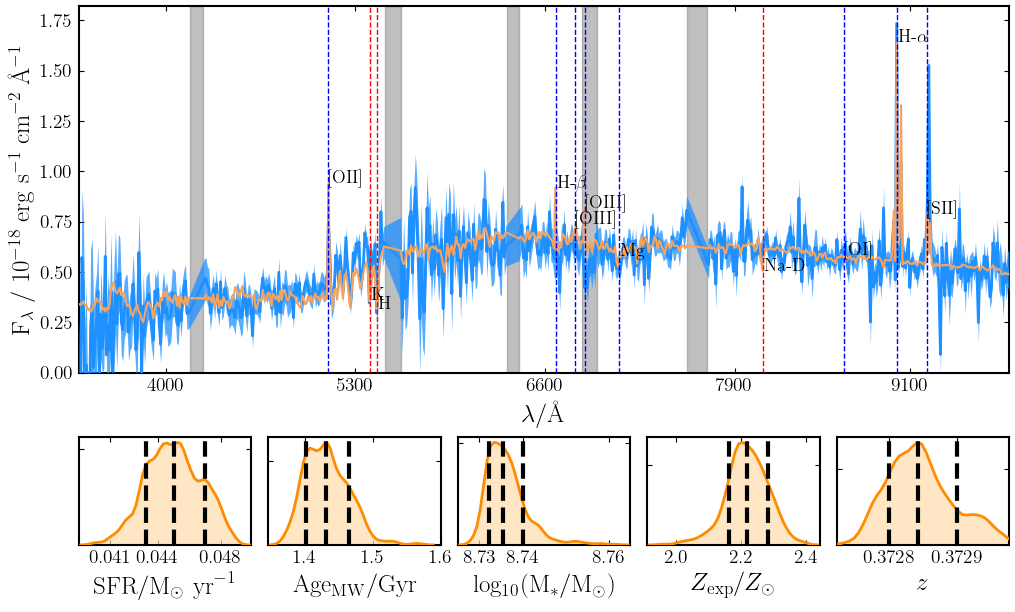}
         \subcaption{}
         \label{bagpipes:ep240708a_H2}
   \end{subfigure}
    \caption{Continued.}
\end{figure*}

\FloatBarrier
\end{appendix}
\end{document}